\documentclass[sigconf,nonacm]{acmart}
\usepackage{tikz}
\usetikzlibrary{positioning,arrows.meta,decorations.pathreplacing}
\usepackage{booktabs}
\usepackage{amsmath}
\usepackage{mathtools}
\usepackage{amsthm}
\usepackage{graphicx}
\usepackage{multirow}
\usepackage{tcolorbox}
\tcbuselibrary{breakable,skins}
\usepackage{caption}
\usepackage{url}
\usepackage{bbm}
\usepackage{placeins}
\usepackage{xspace}
\usepackage{algorithm}
\usepackage{algpseudocode}
\usepackage{pifont} 
\usepackage[textsize=tiny,backgroundcolor=pink]{todonotes}
\usepackage[normalem]{ulem}
\newif\ifshowchanges
\showchangestrue

\newtcolorbox{pbox}{
breakable,
    enhanced,
    pad at break*=1.6mm,
    lines before break=1,
    before upper={\clubpenalty=0 \widowpenalty=0 \interlinepenalty=0 \brokenpenalty=0},
  colback=white,
  colframe=black,
  arc=0mm,
  fonttitle=\bfseries,
  boxrule=0.2mm,
  left=0mm,
  right=0mm,
  top=-0.3mm,
  bottom=-0.3mm
}

\newtcolorbox{probbox}{
breakable,
    enhanced,
    pad at break*=1.5mm,
    lines before break=1,
    before upper={\clubpenalty=0 \widowpenalty=0 \interlinepenalty=0 \brokenpenalty=0},
  colback=red!10,
  colframe=red!10,
  arc=1mm,
  fonttitle=\bfseries,
  boxrule=0.1mm,
  left=0mm,
  right=0mm,
  top=0mm,
  bottom=0mm
}

\newtcolorbox{examplebox}{
breakable,
    enhanced,
    pad at break*=1.5mm,
    lines before break=1,
    before upper={\clubpenalty=0 \widowpenalty=0 \interlinepenalty=0 \brokenpenalty=0},
  colback=red!10,
  colframe=red!20,
  arc=2mm,
  fonttitle=\bfseries,
  boxrule=0mm,
  boxsep=1mm,
  left=0mm,
  right=0mm,
  top=0mm,
  bottom=0mm
}

\theoremstyle{plain}
\newtheorem{theorem}{Theorem}
\newtheorem{lemma}{Lemma}
\newtheorem{proposition}{Proposition}
\newtheorem{corollary}{Corollary}
\newtheorem{example}{Example}
\theoremstyle{definition}
\newtheorem{definition}{Definition}
\newtheorem{problem}{Problem}
\theoremstyle{remark}

\newtheorem{assumption}{Assumption}

\newcommand{\probe}{\textsc{Ledger}\xspace}
\newcommand{\onli}{\textsc{OnliLedger}\xspace}   
\newcommand{\cart}{\textsc{Ledger}\xspace}   
\newcommand{\lexact}{\textsc{Ledger-exact}\xspace}   
\newcommand{\lgreedy}{\textsc{Ledger-greedy}\xspace} 
\newcommand{\lbrute}{\textsc{Ledger-brute}\xspace}   
\newcommand{\lfptas}{\textsc{Ledger-fptas}\xspace}   
\newcommand{\lstate}{\textsc{AnyLedger}\xspace}
\newcommand{\cube}{\textsc{Ball}-tree\xspace}
\newcommand{\oprobe}{\onli}   
\newcommand{\bprobe}{\cart}   

\newcommand{\X}{\mathcal{X}}
\newcommand{\Nstar}{\mathcal{N}^{*}_{k}}
\newcommand{\Nk}{\mathcal{N}_{k}}
\newcommand{\Rec}{\textsc{rec}}

\newcommand{\Cost}{\mathtt{cost}}
\newcommand{\cpr}{c_{\mathrm{probe}}}
\newcommand{\caud}{c_{\mathrm{aud}}}
\newcommand{\cex}{c_{\mathrm{ex}}}
\newcommand{\spend}{\mathtt{spend}}
\newcommand{\Search}{\textsc{Search}}
\newcommand{\dF}{d^{F}} 

\newcommand{\hatC}{\widehat{C}}

\newcommand{\E}{\mathbb{E}}
\newcommand{\Prob}{\mathbb{P}}
\newcommand{\PP}{\mathcal{D}}
\newcommand{\resrate}{\rho}               
\newcommand{\floorrate}{\beta}            
\newcommand{\floorbound}{\bar f}          
\newcommand{\stepsize}{\eta}              
\newcommand{\arate}{\theta}               
\newcommand{\Brem}{B^{\mathrm{rem}}}      
\newcommand{\Rtrue}{R^{\star}}            
\newcommand{\OPT}{\mathrm{OPT}}           
\newcommand{\resgap}{\Delta_{\mathrm{res}}} 
\newcommand{\Rraw}{\widetilde{R}}         
\newcommand{\Ctrue}{C^{\star}}            
\newcommand{\caudmax}{\bar c_{\mathrm{aud}}} 
\newcommand{\lammax}{\lambda_{\max}}      
\DeclareMathOperator*{\argmax}{arg\,max}
\DeclareMathOperator*{\argmin}{arg\,min}

\DeclareMathOperator{\round}{round}

\newcommand{\Qp}{Q_{\mathrm{p}}}   
\newcommand{\Qa}{Q_{\mathrm{a}}}

\usepackage{tikz}\usetikzlibrary{arrows.meta}
\definecolor{figgreen}{RGB}{56,118,29}
\definecolor{figteal}{RGB}{56,142,139}

\begin{document}
\title{\textsc{Ledger}: Learned Budget Allocation For Approximate Nearest Neighbor Search}

\author{Raghav Mittal}
\affiliation{%
  \institution{University Of Texas At Arlington}
  \city{Arlington}
  \country{Texas, USA}
}
\email{rxm0006@mavs.uta.edu}

\author{Minghao Li}
\affiliation{%
  \institution{University Of Toronto}
  \city{Toronto}
  \country{Canada}
}
\email{mingh.li@mail.utoronto.ca}

\author{Nick Koudas}
\affiliation{%
  \institution{University Of Toronto}
  \city{Toronto}
  \country{Canada}
}
\email{koudas@cs.toronto.edu}

\author{Gautam Das}
\affiliation{%
  \institution{University Of Texas At Arlington}
  \city{Arlington}
  \country{Texas, USA}
}
\email{gdas@uta.edu}
\begin{abstract}
Approximate nearest neighbor (ANN) search is increasingly used to power retrieval-augmented generation (a.k.a. \textsc{Rag}) pipelines, which are used to ground LLMs in external data. To keep retrieval accurate, existing methods set the search effort of a query before it runs, with the goal of hitting a predefined \lq recall target\rq\, \textit{at any cost}. For a growing class of compute- or time-constrained workloads, this is not possible. Instead, their goal is to maximize recall under a fixed budget. Recall gains are known to diminish with search effort, and spending a budget uniformly across a query workload either (i) overspends effort on easy queries whose recall has flattened, or (ii) underspends it on hard queries whose recall is climbing. Adaptive methods, meanwhile, remain tied to recall targets. Either way, downstream tasks that operate under tight budgets suffer.
\newline\indent This paper presents the \underline{L}earned \underline{E}ffort Bu\underline{DGE}ting for \underline{R}ecall (a.k.a. $\textsc{Ledger}$) framework, which maximizes the total recall of a workload under a fixed budget of effort. Given a batch of queries, \probe first executes a \lq probe\rq\ up to the sparse upper layer of a hierarchical graph (e.g., HNSW), where a sample of features tells us how hard a query is. \probe then groups the queries with similar hardness feature points into hierarchical spherical clusters that are indexed by the \underline{B}udget-\underline{All}ocation tree, a.k.a. \textsc{Ball}-tree. The \textsc{Ball}-tree uses tables to record cost-recall growth across varying search efforts. With these tables, we show that our problem can be seen as an instance of the multiple-choice knapsack problem, which \probe solves to allocate budgets across a mix of easy and hard queries in the workload. \probe provides guarantees on the quality of budgets given and replaces both (a) fixed-effort searches and (b) recall-tuning methods (that are budget-agnostic) with fast and principled allocations.
\end{abstract}
\maketitle
\pagestyle{standardpagestyle}
\section{Introduction}\label{sec:intro}

Approximate nearest neighbor (ANN) search over learned embeddings is increasingly used to power the retrieval layer of retrieval-augmented generation (a.k.a. \textsc{Rag}), which grounds language models in knowledge fetched from an external corpus~\cite{lewis2020rag}. In standard use, these workloads run on proximity graph indexes such as \textsc{HNSW} and \textsc{ACORN}~\cite{malkov2020hnsw,patel2024acorn}, which connect each vector to a handful of its neighbors, and answer nearest neighbor queries by executing a best-first walk that maintains a bounded set of candidates. The size of that set, a.k.a. the \lq beam width\rq, is the \underline{\textbf{price}} of a query, which refers to the amount of search effort the index commits to it before the search runs. What this effort actually consumes, can be simply measured in terms of time, distance computations, memory, and the like, which we refer to as the \underline{\textbf{cost}} incurred by a query. In graph-based methods, the price of a query is set before the search, while a cost is incurred during the search, and the same price may cost different queries different values. The share of the true nearest neighbors of a query that the walk returns is its recall. Through a set price, every graph index trades cost for recall, which existing literature reports as a curve of recall against queries per second~\cite{aumuller2020annbench}. Meeting a predefined recall target with guarantees is a well-studied problem, and existing systems work to meet it \underline{at any cost}~\cite{li2026spanner,adaef,adaptivebeam}.

\vspace{-3mm}
\begin{figure}[H]
    \centering    \includegraphics[width=\linewidth]{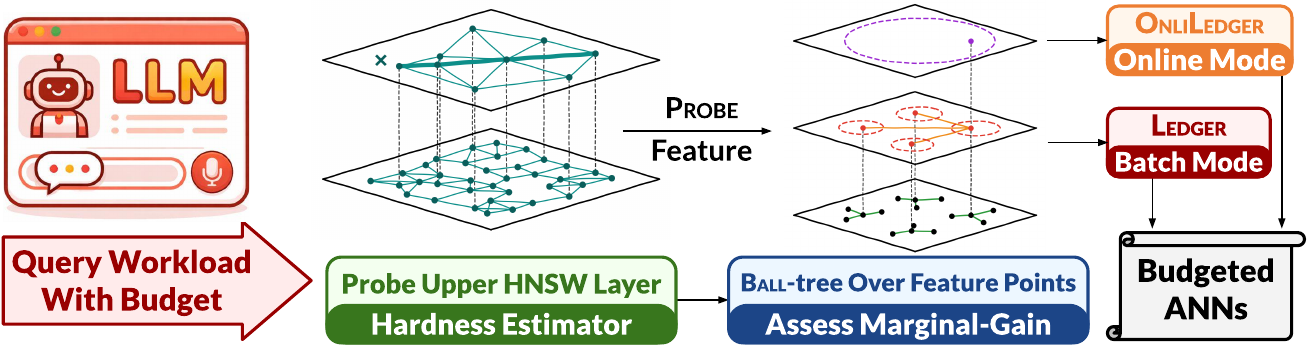}
    \caption{Overview Of The $\textsc{Ledger}$ Framework}
    \label{fig:figone}
\end{figure}
\vspace{-3mm}

\underline{\textbf{Overview Of Adaptive Search:}} graph indexes with fixed beams ignore a well-known and established fact, that queries in a workload may differ significantly in how much search effort they need. ANNs of an easy query may form a tight, far-away cluster that a narrow beam finds easily, while those of a hard query may sit in a dense and crowded region where many candidates look alike, and the beam must increase by a lot to reach them~\cite{learnedtermination3,ceccarello2025workloads}. Adapting effort per query has thus been the focus of existing work. The methods in~\cite{learnedtermination3,tao} learn to predict - from search states or from static features of queries - how much further a search must run to reach a fixed accuracy, while \textsc{Darth} and \textsc{Quake}~\cite{darth2025,quake} take a declared recall target and tune each query’s effort to hit it. \textsc{Auncel}~\cite{auncel} bounds a per-query error instead, while distance-adaptive beam search~\cite{adaptivebeam} grows the beam by relative distances with an accuracy proof. \textit{Whatever their merits, in these papers, a recall target is the input, and the price that must be paid by search is the output.}

\underline{\textbf{The Case For Budgeted Search}}: for a growing class of time- or compute-constrained workloads, the bottleneck is the total cost an index is allowed to incur, which we refer to as its \underline{\textbf{budget}}, and the cost it has incurred so far from that budget is what we refer to as its \underline{\textbf{spend}}. When the budget is fixed in advance and spent uniformly, the workload ends up both (a) wasting effort, on easy queries, for recall it \textit{already had}, and (b) missing, on hard queries, recall it \textit{could have} cheaply and easily afforded when the budget ran out. We argue that budgeted ANN search is a more meaningful formulation, for two reasons. \underline{First}, latency objectives, compute allocations and \textsc{Rag}-timeouts can only be expressed as a search budget, and not as a recall target. \underline{Second}, recall-target algorithms manage latency with a limit on per-query effort, and this rigidity fails where flexibility is in fact most needed. A bad query that is required to exceed the limit is killed and its target is dropped \textit{silently}, while the effort it needed sits unused on easy queries that already met their target earlier. A recall target formulation helps little in describing a budget or flexibly moving effort between queries. To illustrate this, we present the reader with the following example.

\vspace{-2mm}\begin{probbox}
\begin{example}[Agent on a Local Device] Consider a local agentic system running on a laptop or a robot, where compute is limited. It answers nearest neighbor queries from a graph index (such as \textsc{HNSW}~\cite{malkov2020hnsw}). Every vector the search scores costs a distance computation, and the device can afford only a budgeted number of them for the batch of queries the agent issues. The agent has to decide how many distance computations each query of the batch receives.
\end{example}
\end{probbox}
\vspace{-2mm}

Based on the above motivating example, it is intuitive now to understand that an allocation should equalize the \emph{marginal recall per unit of effort} across queries, and not the allocated effort itself. This paper therefore poses and addresses the following question: \textit{given a query workload and a budget, how should the budget be divided across the workload such that its total recall is maximized?}  Note that this idea has a precedent in the database and information retrieval literature on query processing and ranking, where anytime and early-exit algorithms~\cite{mackenzie2022anytime,cambazoglu2010early} are popularly known to exist.

\underline{\textbf{Our Approach}}:  Figure~\ref{fig:figone} depicts the \probe framework, which takes a user-defined budget (e.g., total number of distance computations, memory, execution time, and the like) as input. Our solution works in two steps: (i) a training step that builds the \cube, and (ii) a query step that utilizes it. Specifically, \underline{first}, during training, \probe considers a training dataset of queries (or a uniform sample of the workload, if training data is unavailable), each of which is \lq probed\rq\,(searched) at the base price, i.e., the smallest price of the list (e.g., of candidate beam size values). This probe passes the sparse upper layer that \textsc{HNSW} and other indexes build and stops on the base layer. From the upper layer nearest neighbors we get, \probe reads a set of user-defined hardness features. The values of these features place each query as a point in a feature space, which we refer to as its feature point. Each query of the training dataset is then searched at the base layer at every price (a.k.a. audited), which gives us an estimate of how its recall grows with cost.

The feature points retrieved from these \lq probes\rq\,over the training dataset are indexed by the \cube, where each node (ball) stores the aggregated cost-recall growth across different values of price. The cost-recall growth is (a) \underline{learned} from a handful of audited queries, (b) \underline{pooled} from the nodes above when the audits in a child node are few, and (c) \underline{refined} as more queries in the workload are seen. Each ball answers the question: \textit{for a query whose features land in it, how much more recall does an additional unit of effort buy?}

\underline{Second}, at query time, \probe computes the feature values of each query in the workload by probing it, and based on that, allocates it to a ball in the \cube. Then, it uses a parameter derived from the budget, which we refer to as \underline{\textbf{tariff}}, to address the trade-off between recall and cost. For each candidate price listed in the ball of the query, a table provides an estimated recall that price attains and the cost it incurs. The query is answered at the price that maximizes recall minus tariff times cost. For the online mode, our online method \lq\oprobe\rq\,tracks the tariff based on the cost spent thus far. It raises the tariff when the spend runs ahead of the budget and lowers it when the spend falls behind. For a batched workload, our batched solution uses the probes of every query in the workload before spending the rest of the budget. It reads the cost-recall growth tables learned from the training dataset (or from a uniform sample of the batch). It then spends the budget based on \emph{marginal gain}. \underline{Our core contributions in this paper are multi-fold}:

\textbf{\underline{First}}, we formally state and address the problem of workload-adaptive budgeted approximate nearest-neighbor search on graph indexes, for both streaming as well as batched workloads.
\newline\indent\textbf{\underline{Second}}, we present the \probe framework, which comprises (i) a sparse upper layer probe on HNSW (and other similar indexes), which helps estimate query hardness, (ii) the \cube with balls and aggregated tables that turn a few audited queries into a recall curve over the distribution, and (iii) a tariffed lookup whose spend is monotone in the tariff (as described in Section~\ref{sec:machinery}).
\newline\indent\textbf{\underline{Third}}, we present \oprobe, which solves the online problem by utilizing a subgradient update of the tariff on realized spend. It attains a guaranteed regret against the optimal allocation possible on its tables, and further improves its tables with time (Section~\ref{sec:toll}).
\newline\indent\textbf{\underline{Fourth}}, we present the batched solution, which maps our problem to an instance of the multiple choice knapsack. It first serves the batch at the base price then allocates the rest of the budget exactly, greedily, or by the state of the walks, with guarantees based on known tables and the budgets held on every run (Section~\ref{sec:census}).
\newline\indent\textbf{\underline{Fifth}}, we extensively evaluate \probe on multiple large scale and heterogeneous datasets and workloads. To our knowledge, our paper is the first to allocate a hard budget across a batch or a stream of ANN queries on a graph index. 

\section{Data Model And Problem Statement}\label{sec:model}

We consider a collection $\X=\{x_1,\dots,x_n\}$ of $n$ vectors in $\mathbb{R}^d$ and a distance function on $\mathbb{R}^d$ (smaller meaning closer). A \emph{query} is a pair $(q,k)$ with $q\in\mathbb{R}^d$ and $k\ge1$. We write $q$ alone when $k$ is clear from context. The \emph{exact answer} of a query is a set $\Nstar(q)\subseteq\X$ of $k$ vectors, with ties broken by a fixed rule. A workload $\mathcal{W}$ is either a predefined set or a stream of queries that share a budget $B$.

\textbf{\underline{Inputs}}: we consider any graph index over $\X$ and its search method $\Search(q,k,p)$, where $p$ is the price, or the effort committed to the query before the search runs. This paper considers that $\Search(q,k,p)$ walks the graph in a best-first way and returns the $k$ nearest vectors it has visited, which we denote as $\Nk(q,p)$. The \emph{recall} ($\Rec(q,p)$) of this search is the share of the exact answer that it returns, that is, $\Rec(q,p)=|\Nk(q,p)\cap\Nstar(q)|/k$. This paper also considers that $p$ is drawn from a predefined \emph{price list} $\mathcal{P}=\{p_1<p_2<\dots<p_G\}$ which is finite. For state-of-the-art methods including (but not limited to) \textsc{HNSW}~\cite{malkov2020hnsw}, \textsc{NSG}~\cite{fu2019nsg}, \textsc{DiskANN}~\cite{subramanya2019diskann} as well as \textsc{ACORN}~\cite{patel2024acorn}, we consider the price to be the beam width. However, one may set price as the number of distance computations or the time at which the walk is cut off. As such, \probe is agnostic to the price set by the application.

\begin{definition}[Price And Price List]\label{def:price}
A \emph{price} $p$ is the effort committed to a query before its search runs that the search procedure $\Search(q,k,p)$ takes other than the query. A \emph{price list} $\mathcal{P}=\{p_1<p_2<\dots<p_G\}$ is the set of prices at which a workload may be served, which we consider as input to our problem statements. The smallest price $p_1$ is the \emph{base price}, the cheapest search the index runs, so no price lies below it. Every query is served at it at least, and its search at the base price is its \emph{probe} (Section~\ref{subsec:probe}).
\end{definition}

\begin{definition}[Cost And Budget]\label{def:cost}
A \emph{function} $\Cost(q,p)\ge0$ denotes the actual effort consumed while running $\Search(q,k,p)$ for query $q$ at price $p$. We assume (a) the cost of a set of searches is sum of their costs. (b) $\Cost(q,p)$ can be known only once $\Search(q,k,p)$ has run. (c) $\Cost(q,p)\le\Cost(q,p')$ whenever $p\le p'$. (d) $\Cost(q,p)$ is the cost of the walk on the base layer at price $p$ and that the upper layer probe first costs $\cpr(q)$. A \emph{budget} is a number $B\ge0$ that bounds the maximum cost a query workload may incur.
\end{definition}

\textbf{\underline{Why Allocate}}: the allocated effort at which a query can reach a recall is very well known to vary significantly across a workload~\cite{learnedtermination3,darth2025,wang2024steiner}. A fixed query effort therefore wastes effort on easy queries and misses out on hard queries. It is intuitive that the total recall rises whenever effort moves from a query whose recall has flattened to one whose recall is still rising. The two key problems addressed in this paper (see below) ask for the best such transfer possible within a given query workload. A solution may also choose to \textbf{audit} a query, that is, search it at the top price $p_G$ and then \lq replay\rq\,the search at every lower price (i.e., simulate the search without actually computing distances again) relative to its answer at $p_G$ (Section~\ref{sec:opt}). Its audit cost is $\caud(q)=\cpr(q)+\Cost(q,p_G)$, which correspond to the upper layer descent and the walk to the top price, which contains the walk of its probe. The tables of the later sections work with this relative recall, since an audit can observe it without computing exact neighbors. 


\begin{definition}[Spend]\label{def:spend}
At a point in time, the \emph{spend} on a query refers to what it has consumed from the budget. A query $q$ served at price $p(q)$ spends its probe cost $\cpr(q)$ (Section~\ref{subsec:probe}) plus $\Cost(q,p(q))$, and a query that is audited instead spends its audit cost $\caud(q)$. We write $\spend(q)$ for either, and $\spend_t$ for the $t^{th}$ query $q_t$ in a stream. The spend of a workload $\mathcal{W}$ is simply $\sum_{q\in\mathcal{W}}\spend(q)$.
\end{definition}

\vspace{-1mm}We now have all the necessary ingredients to state the two workload-aware budget allocation ANN problems formally.\vspace{-1mm}

\begin{pbox}
\begin{problem}[Batch \probe Problem]\label{prob:batch}
Given a batched workload $\mathcal{W}$ of $m$ queries and a budget $B$, choose a price $p(q)\in\mathcal{P}$ for every query so as to maximize the total recall within the budget,
\begin{equation}\label{eq:batch}
\max_{p(\cdot)}\ \sum_{q\in \mathcal{W}}\Rec\big(q,p(q)\big)\quad\text{s.t.}\quad\sum_{q\in \mathcal{W}}\spend(q)\le B
\end{equation}
\end{problem}
\end{pbox}

\textbf{\underline{Streaming Case}}: when queries come one by one, a \emph{policy} (denoted $\pi$) must set a price for each one on the spot, without the ability to change the price for a past query and with no knowledge about future queries. The budget is given as the \emph{average budget per query} $b$, and when time horizon $T$ is known the budget is $B=bT$.

\begin{pbox}
\begin{problem}[Online \probe Problem]\label{prob:online}
Given a streaming workload $\mathcal{W}$ of queries $q_1,q_2,\dots,q_T$ under a budget $B=bT$, find the policy $\pi$ that maximizes expected total recall within budget,
\begin{equation}\label{eq:online}
\pi^\star=\argmax_{\pi}\ \E\Big[\sum_{t=1}^{T}\Rec(q_t,p_t)\Big]\quad\text{s.t.}\quad\sum_{t=1}^{T}\spend_t\le B
\end{equation}
\end{problem}
\end{pbox}

A policy is judged by its \emph{regret}, which is the recall it gives up relative to the best allocation (computed by \lexact, Section~\ref{sec:census}).

\begin{figure}[H]
\centering\vspace{-3mm}
\includegraphics[width=\linewidth]{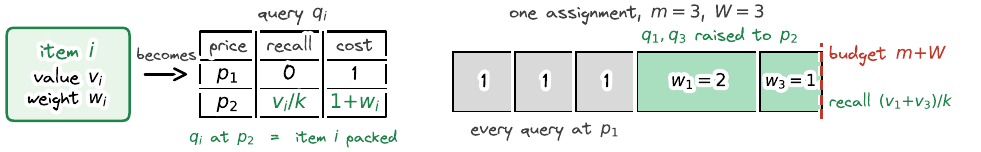}
\caption{Knapsack Items As Two-Price Queries}\vspace{-3mm}
\label{fig:hardness}
\end{figure}

\textbf{\underline{Knapsack Angle}}: consider that we take the budget (e.g., total number of distance computations) as a knapsack and every query as a group of items, one per price, with each weighing the query's cost, and with its worth equal to its recall. If these were known exactly, Problem~\ref{prob:batch} could be trivially mapped to the multiple-choice knapsack, which packs one item of every group (see Figure~\ref{fig:hardness} for intuition). We know that this problem is weakly NP-hard, like the \textsc{0-1 Knapsack}~\cite{garey1979computers}. In this paper, we only work with estimates of them. Since a query's value is at most $k$, a pseudo-polynomial dynamic program solves it exactly with $O(m^2Gk)$ operations.

\section{The $\textsc{Ledger}$ Framework}\label{sec:machinery}

\textbf{\underline{Overview}}: the problem solved by \probe (\underline{L}earned \underline{E}ffort Bu\underline{DGE}ting for \underline{R}ecall) is to estimate, before a query runs, how its recall and cost grow with each price of the price list, which, for a single query, cannot be known beforehand. \probe estimates it from queries that resemble it, in two steps that both our online solution \oprobe and batched solution \bprobe share: (i) a training step that builds the \cube, and (ii) a query step that reads it.

\underline{First}, for training, \probe considers a training dataset of queries (or a uniform sample of the workload queries, if training data is unavailable), which we denote as \emph{probed queries} $\Qp$. Each of them is \emph{probed}, that is, served at the base price ($p_1$), which is the smallest price of the list (Section~\ref{subsec:probe}). This search passes the sparse upper layer of the graph, where it records its $k'$ nearest layer vectors and their distances, and then runs on the base layer at $p_1$. Every method pays this cost, so the probe costs nothing apart from what the cheapest search pays. From its output, a set of user-defined hardness \emph{features} (which \probe is agnostic to) are read, where each feature is a function of what the probe recorded on the layer.

\begin{figure}[H]
\centering
\vspace{-3mm}\includegraphics[width=\linewidth]{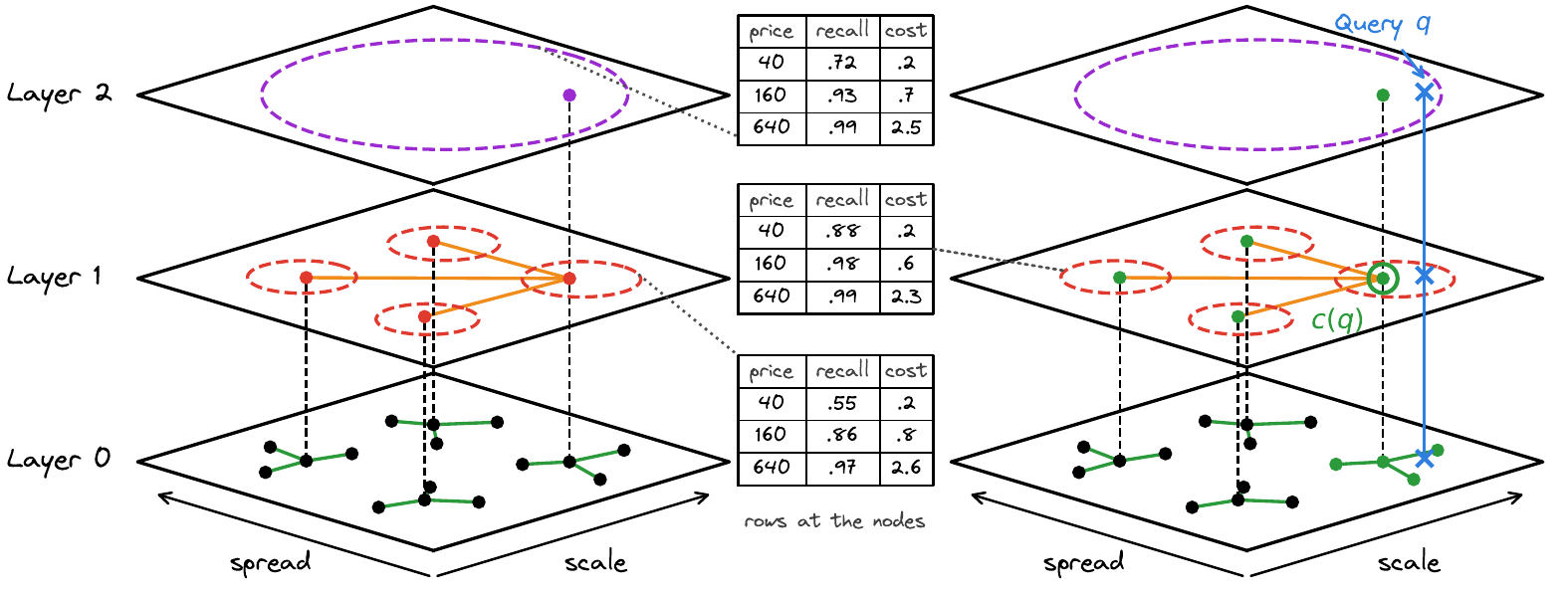}
\caption{The \cube Over Two Features With Its Tables (Left); Placing A Query's Feature Point (Right)}\vspace{-2mm}
\label{fig:cells}
\end{figure}

The values of these features place the query as a point in the \emph{feature space}, which we refer to as its \emph{feature point}. Each query of the training dataset is then \emph{audited}, that is, searched at the base layer at the top price and replayed at every lower price (below), which records its cost-recall growth at every price of the list.
\newline\indent We denote the set of \emph{audited queries} as $\Qa$. For a batch, the tree is built on the batch's own probes and the audits are those of the training dataset. The feature points are organized into the \cube, which is constructed using the principle of hierarchical sampling, in spirit of prior work~\cite{dehghankar2025rar}, so that queries with similar feature points (and correspondingly, similar hardness) fall into the same ball. Every ball of the \cube (as we shall shortly describe) is augmented with tables, which carry the mean recall and the mean cost, at every price $p\in\mathcal{P}$, of the audited queries that fall into it. Higher levels of the tree hold bigger balls, whose tables are averaged over a larger set of feature points. In the \cube, a ball with few audited queries learns its tables from its ancestors in the tree, and the tables are refined as more and more queries are audited.

\underline{Second}, at query time, every new query is probed in the same way, and the cost of its probe, its search at the base price, is spent from the total budget. Its feature point lands in a specific ball (node) of the \cube, and the query inherits the tables of that ball as its estimate. When deciding what price to assign to it, a parameter referred to as the \lq tariff\rq\,turns the tables of the ball into a price (Section~\ref{subsec:tariff}). In essence, it captures the rate at which recall is traded off for cost, and the query is served at the price that maximizes estimated recall minus the tariff times estimated cost. The budget fixes the tariff, which is how both solutions keep within it.

\textbf{\underline{Search Once, Read Every Price}}: a best-first walk at a higher beam visits a superset of the vectors it visits at a lower one, since the beam enters the walk only through its stopping test
, so recall is monotone in the price. Two things can now be said. A search stopped at $p$ can be \emph{resumed} to $p'$ at the marginal cost $\Cost(q,p')-\Cost(q,p)$ instead of restarting, which \bprobe uses to serve a batch in increasing order of price (Section~\ref{sec:census}). And a search at top price $p_G$ contains the search at every lower price, so an \emph{audit} runs a query once at $p_G$, records the walk, and \emph{replays} it to read the recall at every lower price relative to the answer at $p_G$.

\subsection{The Sparse Upper Layer Probe}\label{subsec:probe}

The \emph{probe} of a query is its search at the base price ($p_1$), the smallest price of the list. Every query is served at it at least, under every method, so it is basically free. The search descends the sparse navigation layers of the graph. At one of the upper layers (subject to the graph density), which we refer to as the \emph{probe layer} $L\ge1$, which is a sample $V_L\subset\X$ of $n_L\approx n/M^{L}$ vectors, it records the $k'$ nearest vectors it has visited with distances $D[0]\le\dots\le D[k'-1]$, with
\begin{equation}\label{eq:kprime}
k'=\min\big\{\max\{\round(k\,n_L/n),\,k'_{\min}\},\,n_L\big\}.
\end{equation}
The ball around $q$ that holds its $k$ nearest base vectors holds about $k\,n_L/n$ layer vectors, so where this count reaches $k'$ the $k'^{th}$ nearest layer vector sits at about the distance of the $k^{th}$ nearest base vector, and the minimum $k'_{\min}$ keeps enough layer vectors to read at a small $k$ or on a sparse layer. These vectors seed its walk on the base layer, which runs at $p_1$ and then pauses, with its candidate heap and its visited set kept, so that a higher price resumes it instead of restarting. The output of the probe is recorded on the layer, which the features read, and the state of its paused walk,  which our anytime allocation reads (as we shall shortly describe in Section~\ref{subsec:state}). The probe layer is the user’s choice, and M guides it. A higher layer makes the descent cheaper and its records coarser. At a large M every vertex has many links, so a higher layer still gives the probe enough vectors to read. On a sparser graph the probe records on layer one, and may go on into a part of the base layer. Our experiments, at M of 16 and 32, execute the probe on layer two. Note that \textsc{HNSW} has a sparse upper layer by construction, and a flat graph index can get one by sampling $n/M$ vectors and linking them as the base layer is linked. In standard use, the upper layers stay in memory while the base layer stays on disk, so the upper layer part of the probe is cheap. The cost of the probe is $\cpr(q)+\Cost(q,p_1)$, the descent through the upper layers and the walk at the base price, and it is spent from the total budget like that of any search.

\subsection{\cube: Feature-Agnostic Construction}
\begin{figure}[H]
\centering
\vspace{-2mm}\includegraphics[width=\linewidth]{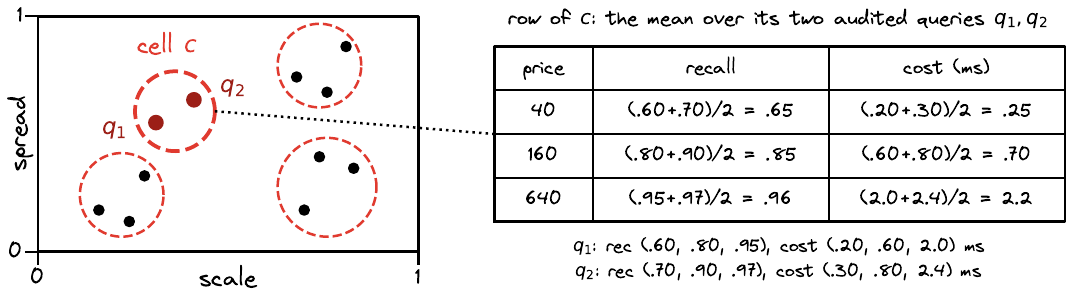}
\caption{Example: One Ball With Two Audited Queries And Its Table, Computed Entry By Entry For Illustration}\vspace{-2mm}
\label{fig:cellexample}
\end{figure}

The \cube is built in three steps. \underline{First}, every query of the training dataset is probed, that is, served at the base price (Section~\ref{subsec:probe}), in the same way as a query at run time. \underline{Second}, a set of user-defined hardness features turns their probe output into a point in the feature space, which we refer to as their \emph{feature point}. \underline{Third}, the queries of the training dataset are audited at the base layer, i.e., fully searched at every price of the price list to get their recall and cost. The feature points earlier obtained from the training dataset are then grouped into balls, and every ball is augmented (using a pointer) with a table of the mean recall and mean cost of its audited queries at every price (see Figure~\ref{fig:cellexample} for an example). A query feature point that later lands in a ball at query time reads this average cost-recall growth as its estimate, before its own search runs. The features only need to learn how hard the query is, and existing works propose many heuristic features that can be easily plugged into \probe, including those proposed in \textsc{LAET}~\cite{learnedtermination3}, \textsc{Tao}~\cite{tao}, and \textsc{DARTH}~\cite{darth2025}. \textit{\underline{\textbf{Note}}: As such, we do not restrict ourselves to any one of these. Instead, we propose a broad framework that is easy to adopt. Figure~\ref{fig:cellexample} shows an example of balls and how their augmented tables are computed, by assuming two features, namely spread and scale.}

\textbf{\underline{The Feature Space}}: recall, that the probe of a query $q$ records its $k'$ nearest layer vectors and their distances $D_q=(D[0],\dots,D[k'-1])$. A \emph{feature} is a function $f$ of the layer records alone, so that it is known before the base-layer walk begins and never depends on it, and a \emph{feature map} $\phi=(f_1,\dots,f_{\dF})$, with $\dF$ the number of features, sends a query to the point $\phi(q)\in\mathbb{R}^{\dF}$. The \emph{feature space} $\Phi\subseteq\mathbb{R}^{\dF}$, of dimension $\dF$, is the set of such points over the workload. \cube is built over $\Phi$ alone.Every point of $\Phi$ is placed in exactly one ball, through $\phi(q)$ alone, and adding or swapping a feature adds or swaps a coordinate of the points the tree is built over, while nothing else in the method changes. Table~\ref{tab:features} lists the features used in our experiments, that are all read from what the probe recorded on the layer. The first two are computed in the following manner
\begin{align}
\mathrm{scale}(q) &= D[k'-1], \label{eq:scale}\\
\mathrm{spread}(q) &= D[k'-1]-D[0], \label{eq:spread}
\end{align}
with $\mathrm{scale}=\infty$ and $\mathrm{spread}=0$ when the probe finds fewer than $k'$ vectors. Scale is essentially the radius the search must reach, and the price a query needs tends to rise with it. Spread measures whether the nearest neighbor stands out or whether all $k'$ neighbors sit in one small cluster, where the beam may have many near-equal candidates, so the price falls with spread. The others are defined in Table~\ref{tab:features}, with the hardness measures of the literature among them read on the probe instead of on the true neighbors. 

\subsection{Hierarchical Sampling Of Feature Points}\label{subsec:cells}

The key idea behind the \cube is that queries with similar feature points behave alike. For instance, for features spread and scale, we assume that two queries whose neighbors are equally far and spread, will show a similar cost-recall growth at every price. With this insight, \probe clusters beforehand the feature points of the probed training dataset (by the sampling described below), and every later query that lands in that cluster inherits its estimates. In \probe, each feature is first rescaled to $[0,1]$ by its rank among the probed queries of the training dataset, so that no feature dominates the distance between two points ($\phi(q)\in[0,1]^{\dF}$).

\textbf{\underline{$\varepsilon$-Sampling On Feature Points}}: the \cube is built over the training dataset, which must preserve how the feature points of the workload spread over the feature space. We state this using the concept of the $\varepsilon$-samples, in spirit of prior literature~\cite{dehghankar2025rar}.

\begin{definition}[$\varepsilon$-Sample]\label{def:epssample}
The probed queries $\Qp$ are an \emph{$\varepsilon$-sample} of the workload $\mathcal{W}$ if, for every ball $A$ of the feature space, the proportion of queries whose feature points lie in $A$ is approximately preserved in $\Qp$, that is,
\begin{equation}\label{eq:epssample}
\left|\frac{|\{q\in\mathcal{W}:\phi(q)\in A\}|}{|\mathcal{W}|}-\frac{|\{q'\in\Qp:\phi(q')\in A\}|}{|\Qp|}\right|\le\varepsilon .
\end{equation}
\end{definition}

The VC dimension~\cite{vapnik1971uniform} of the balls is the size of the largest set of points of which every subset is exactly the part that lies inside some ball. Balls in $\mathbb{R}^{\dF}$ have VC dimension $\dF+1$~\cite{dudley1979balls}, which gives the following theorem.

\begin{pbox}
\begin{theorem}[$\varepsilon$-Sample Theorem~\cite{vapnik1971uniform}]\label{thm:epssample}
For any $\varepsilon,\delta\in(0,1)$, a random sample of the workload of size $O\big(\frac{\dF+1}{\varepsilon^2}\log\frac{\dF+1}{\varepsilon}+\frac{1}{\varepsilon^2}\log\frac{1}{\delta}\big)$ is an $\varepsilon$-sample of it with probability at least $1-\delta$.
\end{theorem}
\end{pbox}

The theorem also holds for a sample drawn from the distribution of the workload, as a training dataset is~\cite{vapnik1971uniform}. So a training dataset of that size (or a uniform sample of the workload, if training data is unavailable) is an $\varepsilon$-sample with probability at least $1-\delta$, and the size grows only with the number of features.

\textbf{\underline{Building The \cube}}: the tree is built over the feature points of $\Qp$ by hierarchical sampling, in the spirit of prior literature~\cite{dehghankar2025rar} that builds such a hierarchy over the vector database itself, with its \emph{decay rate} $r\ge2$ as a parameter. In the \cube, layer $0$ is all of $\Qp$, $\Qp^{(0)}=\Qp$, and every layer $\Qp^{(\ell+1)}$ is a uniform random subset of $\Qp^{(\ell)}$ that is $r$ times smaller, $|\Qp^{(\ell+1)}|=\lceil|\Qp^{(\ell)}|/r\rceil$, until a layer holds only a single point (i.e., the root). Every point $v$ of a lower layer is linked to its nearest point of the layer above, that is, its parent. A \emph{node} is a point of a layer, its \emph{region} is the set of points of $\Qp$ that reach it through parent links, and its \emph{ball} is the smallest sphere that encloses the feature points of its region. In the \cube, regions are nested, and so the ball of a node contains every probed query below it. A query $q$ is placed in the ball of layer $1$ whose point is nearest to its feature point. We call it the \emph{ball of $q$}, denoted as $c(q)$,
\begin{equation}\label{eq:place}
c(q)=\argmin_{c'\in\Qp^{(1)}}\ \|\phi(q)-\phi(c')\|_2,
\end{equation}
which is found by descending from the root and skipping every node whose ball is farther from $\phi(q)$ than the nearest point of layer $1$ found so far (Figure~\ref{fig:cells}, right). A query whose feature point lies outside the ball of $c(q)$ resembles no probed query. It still reads the row of $c(q)$, which our pooling method below takes from the ancestors. The \emph{balls} $\mathcal{B}$ are all nodes above the base layer (about $|\Qp|/(r-1)$ in total) and every ball in \cube is augmented with a table. The base layer, whose nodes are single queries, are only there to form the regions and balls and is not kept. Once built, the \cube is static and not changed until it is rebuilt for a drifted workload, so between rebuilds $c(q)$ is a function of $q$ alone.

\begin{algorithm}[t]\small
\caption{\textsc{Build-\cube}$(\Qp,\Qa,r)$: the tree and its tables}\label{alg:balltree}
\begin{algorithmic}[1]
\Require probed queries $\Qp$, audited queries $\Qa$, feature map $\phi$, price list $\mathcal{P}$, decay rate $r$, cap $m_{\max}$
\setcounter{ALG@line}{0}\State $\Qp^{(0)}\gets\Qp$;\quad $\ell\gets0$ \Comment{base layer}
\State $\mathrm{reg}(q)\gets\{q\},\ \forall q\in\Qp$ \Comment{regions}
\While{$|\Qp^{(\ell)}|>1$} \Comment{$\log_r|\Qp|$ layers}
  \State $\Qp^{(\ell+1)}\gets\mathrm{sample}\big(\Qp^{(\ell)},\lceil|\Qp^{(\ell)}|/r\rceil\big)$ \Comment{uniform}
  \State $\mathrm{par}(v)\gets\argmin_{u\in\Qp^{(\ell+1)}}\|\phi(v)-\phi(u)\|_2,\ \forall v\in\Qp^{(\ell)}$
  \State $\mathrm{reg}(u)\gets\bigcup_{v:\,\mathrm{par}(v)=u}\mathrm{reg}(v),\ \forall u\in\Qp^{(\ell+1)}$
  \State $\mathrm{ball}(u)\gets\textsc{MinSphere}\big(\{\phi(q)\}_{q\in\mathrm{reg}(u)}\big),\ \forall u\in\Qp^{(\ell+1)}$
  \State $\ell\gets\ell+1$
\EndWhile
\State $\mathcal{B}\gets\bigsqcup_{j=1}^{\ell}\Qp^{(j)}$ \Comment{balls}
\State $c(q)\gets\argmin_{c'\in\Qp^{(1)}}\|\phi(q)-\phi(c')\|_2,\ \forall q\in\Qa$ \Comment{Eq.~\eqref{eq:place}}
\State $\Qa(c)\gets\{q\in\Qa:c(q)\in\mathrm{reg}(c)\}$;\quad $m_c\gets|\Qa(c)|,\ \forall c\in\mathcal{B}$
\State $(\bar R_c,\bar C_c)\gets\frac{1}{m_c}\sum_{q\in\Qa(c)}\big(\Rec(q,\cdot),\Cost(q,\cdot)\big),\ \forall c\in\mathcal{B}$ \Comment{Eq.~\eqref{eq:tables}}
\State $(\Rraw_c,\widetilde C_c)\gets(\bar R_c,\bar C_c),\ \forall c\in\Qp^{(\ell)}$ \Comment{root}
\For{$j=\ell-1,\dots,1$} \Comment{top-down}
  \State $m_0\gets\min\{m_{\max},\hat\sigma^2/\hat\tau^2\}$ 
  \State $(\Rraw_c,\widetilde C_c)\gets\frac{m_c(\bar R_c,\bar C_c)+m_0(\Rraw_{\mathrm{par}(c)},\widetilde C_{\mathrm{par}(c)})}{m_c+m_0},\ \forall c\in\Qp^{(j)}$ \Comment{Eq.~\eqref{eq:pool}}
\EndFor
\State $\hatC_c\gets\textsc{Isotonic}(\widetilde C_c)$;\quad $R_c\gets\textsc{Majorant}(\hatC_c,\Rraw_c),\ \forall c\in\mathcal{B}$
\State $\gamma_c(g)\gets\frac{R_c(p_{g+1})-R_c(p_g)}{\hatC_c(p_{g+1})-\hatC_c(p_g)},\ \forall c\in\mathcal{B},\ g<G$ \Comment{Eq.~\eqref{eq:density}}
\State \Return $\{(R_c,\hatC_c)\}_{c\in\mathcal{B}}$
\end{algorithmic}
\end{algorithm}

\textbf{\underline{What Balls Give}}: layer $\ell$ has $|\Qp|/r^{\ell}$ nodes (balls), and each of them holds $r^{\ell}$ probed queries on average, so the balls of layer 1 are the smallest, with $\sim|\Qp|/r$ of them irrespective of what the number of features is. Their tables are learned and pooled with those of their parent balls (Section~\ref{subsec:tables}). If a ball has very few audits, it inherits the rows of its ancestors and detaches from them as and when its own audits arrive. By doing so, the resolution at which a query is read is set by the audits themselves, which are finer in areas where they are dense and coarser where they are sparse, and \cube automatically decides which ball's tables the query needs.

\vspace{-1mm}\begin{pbox}
\begin{proposition}[Representativeness]\label{prop:epsball}
Let $\Qp$ be an $\varepsilon$-sample of the workload $\mathcal{W}$. Then every ball of the tree carries a share of $\mathcal{W}$ within $\varepsilon$ of its share of $\Qp$, and the region outside root's ball, where a query resembles no probed query, carries at most an $\varepsilon$ share of it.
\end{proposition}
\end{pbox}\vspace{-1mm}

Note that the \cube also works for a dynamic and drifting workload. Specifically, when a new probed query enters $\Qp$, it is promoted through the layers with probability $1/r$ each, and is linked to its nearest point above. When many queries fall outside the ball of their node, the tree is rebuilt from $\Qp$ and the tables from $\Qa$. 

\textbf{\underline{Why A Sampled Hierarchy}}: in principle, a quantile grid constructed over the hardness features would also work with \probe, but its buckets multiply with every feature added. A kd-tree~\cite{bentley1975multidimensional} splits one feature at a time, so its cells are boxes set by the order of the splits, and a query near a face of its box reads a cell that its nearest probed queries do not share. Here the groups are nested by construction, and as we have said, their number is set by $r$ alone whatever the number of features. A ball is set by the points it holds and not by the axes. The hierarchy follows~\cite{dehghankar2025rar}, where it is built over the vectors themselves and, on range queries, keeps pruning at dimensions where kd-trees and R-trees stop. 

\subsection{Maintaining The \cube Under Drift}\label{app:dynamic}
Remember the \cube is built once over the probed queries $\Qp$, which is what makes $c(q)$ a fixed function of $q$. To maintain the \cube under drift, we perform two operations. \underline{\emph{Insert}}: a new audited query enters the base layer and is promoted through the layers with probability $1/r$ each (one Bernoulli draw per layer~\cite{vitter1985random}). Correspondingly, it is linked to its nearest point of the layer above, and the spheres of its parents node grow to include it. \underline{\emph{Delete}}: the oldest audit beyond the window needs to be deleted. A node it owns keeps its sphere until the next rebuild, and when the live points of a layer fall below half the size it had at its last build, the layers are rebuilt from the live points. A \lq\emph{monitor}\rq reads this tree. For a new batch, the monitor evaluates the share of the batch queries whose feature points lie outside the sphere of their node. If a batch exceeds $1.5$ times that level it is flagged, and we perform audits on a uniform sample of its queries after serving them. The audits join $\Qa$ and the tree, and the oldest audits are asked to leave beyond a predefined window. Then the tree is rebuilt from the live points.

\subsection{Attaching Cost-Recall Tables To \cube}\label{subsec:tables}

\textbf{\underline{Motivation}}: the benefit of constructing balls over the feature points of the training dataset is that it tells us how the cost and recall of its queries grow with price (e.g., beam width). In \probe, every query of the training dataset that is audited (denoted $\Qa$) is searched at the top price and replayed at every price $p\in\mathcal{P}$, so its recall relative to the top price (Section~\ref{sec:model}) and its cost at every price of the list are known. For a ball $c\in\mathcal{B}$ with $m_c$ audits (the audited queries $\Qa(c)$ in $c$ or below it), the raw tables are its mean recall ($\bar R_c(p)$) and mean cost ($\bar C_c(p)$) at every price, which we simply denote as:

\begin{equation}\label{eq:tables}
\begin{gathered}\vspace{-1mm}
\bar R_c(p)=\frac{1}{m_c}\!\!\sum_{q\in\Qa(c)}\!\!\Rec(q,p),\\
\bar C_c(p)=\frac{1}{m_c}\!\!\sum_{q\in\Qa(c)}\!\!\Cost(q,p).
\end{gathered}\vspace{-1mm}
\end{equation}

Note that in a stream the mean cost values additionally count \textit{every query of the ball served at price $p$}, irrespective of whether they are audited or not, since cost is actually calculated for every search, while recall cannot be known, and so it must wait for a full audit.

\textbf{\underline{\lq Learning\rq\,The Tables}}: in the \cube, balls with few audits have noisy and inaccurate tables. Further, in the online case, for early queries in the stream, most balls have no tables at all, and so, the tables must be learned. We do this in two steps. The first step \emph{pools} the tables down from the root node to the leaves. A ball's own row is weighted by its $m_c$ audits. Its parent's row counts as if it were $m_0$ more audits, for a given pseudo-count $m_0\ge0$. A ball with no audits therefore only reads its parent's row. A ball with $m_c\gg m_0$ audits reads its own. So $m_0$ sets how many audits a ball needs before its own evidence outweighs its parent's. At the root the pooled row is the raw row, and at every other ball $c$ in a top-down order:
\begin{equation}\label{eq:pool}
\Rraw_c=\frac{m_c\,\bar R_c+m_0\,\Rraw_{\mathrm{par}(c)}}{m_c+m_0},
\end{equation}
where $\Rraw_c$ is the pooled row. The cost tables are pooled in the same way. We set $m_0=\sigma^2/\tau^2$, where $\sigma^2$ is the variance of one audited recall value within a ball, and $\tau^2$ is the mean squared difference between a ball's true row and its parent's true row. This ratio
sets the shrinkage weight of empirical Bayes,
as in the James-Stein estimator~\cite{efron1973stein,mccallum1998shrinkage}. Both variances are measured from existing audits. The pooled variance within the balls of a depth gives $\hat\sigma^2$. Over the $J$ balls $c$ of that depth that hold an audit,
\begin{align*}
\hat\tau^2=\max\Big\{0,\ \tfrac1J\textstyle\sum_{c}\big(\bar R_c-\bar R_{\mathrm{par}(c)}\big)^2-\tfrac1J\sum_{c}\hat\sigma^2\big(1/m_c-1/m_{\mathrm{par}(c)}\big)\Big\}
\end{align*}
averaged over the prices below $p_G$, with $m_0$ capped at $m_{\max}$. The second step makes the pooled row concave (below) so the guarantee of the batch greedy follows.

\begin{figure}[H]
\centering\vspace{-3mm}
\includegraphics[width=0.96\linewidth]{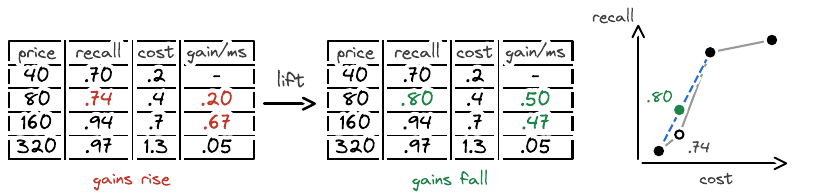}
\caption{Lifting Dented Entry Makes Gains Nonincreasing}\vspace{-3mm}
\label{fig:concave}
\end{figure}

\textbf{\underline{Making Rows Concave}}: allocating budgets based on marginal gain needs rows in which every extra unit of effort buys less recall than the one before it. A noisy row may break this. In Figure~\ref{fig:concave} the step to price $80$ buys $0.20$ recall per ms and the next step $0.67$, so a greedy stops at the poor step and never reaches the good one. We fix it by raising the dented entry to the straight line between its neighbors ($0.74$ to $0.80$), after which the gains become monotonic. Such a row is \emph{cost-concave}. The final rows $(R_c,\hatC_c)$ of all balls are their \emph{tables}, and a query $q$ reads the row of its ball $c(q)$.

\subsection{Tariff: The Cost-Recall Tradeoff}\label{subsec:tariff}

In our setting, note that cost and recall are inherently at odds, since higher recall can only be bought at higher cost. To model this tradeoff, we introduce a \emph{tariff} $\lambda \ge 0$, which is measured in units of recall per unit of cost, and fixes the rate at which one is exchanged for the other. Under tariff $\lambda$, a query $q$ is served at the \emph{tariffed price}
\begin{equation}\label{eq:tariff}
p^\lambda(q)=\argmax_{p\in \mathcal{P}}\ \big\{\Rraw_{c(q)}(p)-\lambda\,\hatC_{c(q)}(p)\big\},
\end{equation}
with ties going to the lower price. The row $\Rraw_c$ in Equation~\eqref{eq:tariff} is the pooled row before the concave lift of Section~\ref{subsec:tables}, the row whose error the online analysis bounds. The lifted row serves the batch greedy alone. A higher tariff makes cost weigh more and moves every ball to a lower price, which is how a budget is met (as we shall shortly discuss). Based on the tariff, a ball whose cost-recall growth has flattened stops at a cheap price and a ball whose row is still rising keeps going, so every served query stops where its marginal recall per unit of cost has dropped to $\lambda$. Two properties of the tariff carry the guarantees of the next two sections.

\begin{pbox}
\begin{lemma}\label{lem:tariff}
For a set of tables, (a)  predicted cost of every query is nonincreasing in $\lambda$, and (b) for any $\lambda\ge0$ and any workload $\mathcal{W}$, allocations whose predicted cost does not exceed the tariff's has predicted recall no bigger than tariff's~\cite{everett1963}.
\end{lemma}\vspace{-2mm}
\end{pbox}
\section{$\onli$: The Online Solution}\label{sec:toll}

\label{op:overview}
\oprobe serves each query of the stream at a tariffed price (based on Equation~\eqref{eq:tariff}) and adjusts the tariff after every query so that the spend tracks the budget rate. The tariff update strategy used in this work is in the spirit of prior literature~\cite{balseiro2023dual}, which discusses the concepts of random rewards and costs and per-query decision errors. In the context of nearest neighbor search, we point out two things: (a) in our setting, every query must be answered at least at the base price, so there is no zero-cost action, which~\cite{balseiro2023dual} assume, and (b) the recall of each price is learned from audits paid from the same budget. Sections~\ref{subsec:op:floor} and~\ref{subsec:op:audits} handle each, and Section~\ref{subsec:op:guarantees} separates the approximation loss of the balls and the price list from the overhead of learning the tables.

\subsection{The Tariff Update}\label{subsec:op:update}
\oprobe initializes the tariff $\lambda_1\in[0,\lammax]$ on the training dataset and, after serving $q_t$, takes a projected subgradient step on the realized spend~\cite{balseiro2023dual}, which we simply write as
\begin{equation}\label{eq:tariffupdate}
\lambda_{t+1}=\min\big\{\lammax,\ \max\{0,\ \lambda_t+\stepsize\,(\spend_t-b)\}\big\},
\end{equation}
where $\spend_t$ is the spend of $q_t$, descent included (Definition~\ref{def:spend}). The cap $\lammax\ge1/(b-\floorbound)$ bounds the tariff (Assumption~\ref{asm:feasible}). Overspending raises the tariff and moves later queries to lower prices.

\subsection{The Base Price And The Reserve}\label{subsec:op:floor}
\label{op:floor}Every query is served at least at the base price $p_1$, and the tariff prices only the upgrade above it. Recall and cost are measured above the base price, as $\Rraw_c(p)-\Rraw_c(p_1)$ and $\hatC_c(p)-\hatC_c(p_1)$, which makes $p_1$ the zero-cost action of the upgrade and leaves the tariffed price unchanged. As a safeguard, the search at the base price stops before an expansion (of cost at most $\cex$) that could raise the base spend $\cpr(q)+\Cost(q,p_1)$ above a bound $\floorbound$, and the query is then answered with the $k$ nearest vectors visited. Under Assumption~\ref{asm:feasible} the cap never binds, so $\Rec(q,p)$ and $\Cost(q,p)$ always denote the uncapped search. The \emph{residual rate} $\resrate=b-\floorrate$, with $\floorrate=\E_{q\sim\PP}[\cpr(q)+\Cost(q,p_1)]$ the mean base spend and $\PP$ the distribution of the stream, from which its queries are drawn independently, is the budget per query available for upgrades.

\begin{pbox}
\begin{assumption}[Feasible Budget]\label{asm:feasible}
Every base spend $\cpr(q)+\Cost(q,p_1)$ is at most $\floorbound-\cex$, and $\floorbound<b$.
\end{assumption}
\end{pbox}
\noindent The assumption implies $\floorrate\le\floorbound$, so $\resrate\ge b-\floorbound>0$, and that every stream can be served in full at $p_1$.

\textbf{\underline{The Reserve Check}}:\label{op:reserve}
to keep the budget on every stream, not only in expectation, \oprobe holds a reserve of $\floorbound$ for every query still to come. With $\Brem_t=B-\sum_{s<t}\spend_s$, the search of $q_t$ always completes the base price $p_1$, and above it starts an expansion only if $\Brem_t-\spend_t-\cex\ge(T-t)\floorbound$ (line~\ref{ln:op:expand} of Algorithm~\ref{alg:onliprobe}; $\cex\le2M$ when cost counts distance computations, and $3M$ with the 8-bit code test).

\begin{pbox}
\begin{proposition}[Budget]\label{prop:budget}
Under Assumption~\ref{asm:feasible}, the total spend of \oprobe is at most $B$ on every stream.
\end{proposition}
\end{pbox}

\begin{examplebox}\begin{example}[Worked Tariff Example]\label{ex:tariff}\small
Consider a price list with three ef\_search values $p_1<p_2<p_3$, at which every query spends $1$, $2$ and $4$ units (say, a unit is a thousand distance computations). A query in an \emph{easy} ball (E) has recall $0.90$, $0.94$ and $0.95$ at these ef\_search values, and a query in a \emph{hard} ball (H) has recall values $0.50$, $0.80$ and $0.90$. The stream has $T=8$ queries and a budget of $b=1.5$ units per query ($B=12$, or 12k distance computations). The step is $\stepsize=0.1$, and the tariff starts at $\lambda_1=0.26$.

\vspace{2mm}{\centering\footnotesize\upshape\setlength{\tabcolsep}{3.5pt}
\begin{tabular}{l|cccccccc}
\hline
Query $t$ & $1$ & $2$ & $3$ & $4$ & $5$ & $6$ & $7$ & $8$\\
\hline
Ball & E & H & H & H & E & E & H & E\\
Tariff $\lambda_t$ & $0.26$ & $0.21$ & $0.26$ & $0.31$ & $0.26$ & $0.21$ & $0.16$ & $0.21$\\
Price $p_t$ & $p_1$ & $p_2$ & $p_2$ & $p_1$ & $p_1$ & $p_1$ & $p_2$ & $p_1$\\
Spend & $1$ & $2$ & $2$ & $1$ & $1$ & $1$ & $2$ & $1$\\
\hline
\end{tabular}\par}\vspace{2mm}

\noindent Using Equation~\ref{eq:tariff}, an easy query takes price $p_1$ at any tariff of $0.04$ or more, since its upgrade to $p_2$ buys $0.04$ recall for one unit. A hard query's upgrade to $p_2$ buys $0.30$ for one unit and its upgrade to $p_3$ buys $0.10$ for two, so it takes $p_2$ at a tariff between $0.05$ and $0.30$, and $p_1$ from $0.30$ on. Query~$1$ is easy and spends $1$, which is $0.5$ under the rate $b$, so Equation~\ref{eq:tariffupdate} lowers the tariff by $0.1\cdot0.5=0.05$. Queries~$2$ and~$3$ are hard and take $p_2$. Each spends $0.5$ over the rate, and the tariff climbs to $0.31$. The stream is now ahead of its budget, so the tariff sends the third hard query in a row to $p_1$. The easy queries that follow bring the tariff down again, and the next hard query is upgraded. The eight queries spend $11$ of their budget of $12$ and reach a total recall of $6.5$. A fixed price (ef\_search) for all eight queries would have used $p_1$ to keep cost within this budget, and its total recall would have been lower.
\end{example}\end{examplebox}


\subsection{Audits And Tables}\label{subsec:op:audits}
The tables start from the audited training queries $\Qa$. An audit searches to $p_G$ and records the expansions as a replay audit (Section~\ref{sec:machinery}), which yields the recall and the cost at every price. Its spend $\cpr(q)+\Cost(q,p_G)$ is assumed at most a bound $\caudmax$, set from the training dataset. Query $q_t$ has \emph{slack} when $\Brem_t-(T-t)\floorbound\ge\caudmax+\cex$, where the $\cex$ covers the one-expansion look-ahead of the reserve check, and it is audited when it has slack and an independent coin with probability $\arate$ comes up heads. Both are decided before the probe and depend only on the past, so every audit finishes within the reserve and the audits of a ball are a uniform sample of that ball. Every query with slack also records its cost $\Cost(q_t,p)$ at every price $p\le p^\dagger$, which updates $\hatC_c$ (the descent stays outside the row, as in Equation~\eqref{eq:tables}). A query without slack records nothing, since the reserve may cut it where it is hard. An update re-pools and re-fits only the rows on the path from the ball to the root, while the tree itself is fixed. \oprobe evaluates Equation~\eqref{eq:tariff} on the pooled rows $\Rraw_c$ of Equation~\eqref{eq:pool}, not on the cost-concave lifts of Section~\ref{subsec:tables}, for the reason given in Section~\ref{subsec:op:guarantees}.
Algorithm~\ref{alg:onliprobe} presents our overall approach, where lines~\ref{ln:op:slack}--\ref{ln:op:coin} decide the audit before the probe, line~\ref{ln:op:target} sets the target price, lines~\ref{ln:op:search}--\ref{ln:op:answer} serve, where $p_t$ is the highest price whose search has completed, and lines~\ref{ln:op:cost}--\ref{ln:op:tariff} learn and re-price.

\vspace{-2mm}\begin{algorithm}[H]\small
\caption{\oprobe}\label{alg:onliprobe}
\begin{algorithmic}[1]
\Require price list $\mathcal{P}$, pooled rows $(\Rraw_c,\hatC_c)$, budget $B=bT$, step $\stepsize$, tariff cap $\lammax$, audit rate $\arate$, bounds $\floorbound,\cex,\caudmax$, tariff $\lambda_1$
\State $\Brem\gets B$ \Comment{remaining budget}
\For{$t=1,\dots,T$}
  \State $s_t\gets\big[\Brem-(T-t)\floorbound\ge\caudmax+\cex\big]$ \Comment{slack}\label{ln:op:slack}
  \State $a_t\gets s_t\cdot\mathrm{Bernoulli}(\arate)$ \Comment{audit coin, before probe}\label{ln:op:coin}
  \State descend $q_t$; $c\gets c(q_t)$ \Comment{Sec.~\ref{subsec:probe}}
  \State $p^\dagger\gets p_G$ if $a_t=1$, else $p^{\lambda_t}(q_t)$ \Comment{target price}\label{ln:op:target}
  \While{$p^\dagger$ not reached}\label{ln:op:search}
    \State \textbf{if} $a_t=0$, $p_1$ not reached, $\spend_t+\cex>\floorbound$: \textbf{break} \Comment{safeguard}\label{ln:op:floorcap}
    \State \textbf{if} $p_1$ reached, $\Brem-\spend_t-\cex<(T-t)\floorbound$: \textbf{break}\label{ln:op:reserve}
    \State expand next vertex of $\Search(q_t,k,p^\dagger)$ \Comment{cost $\le\cex$}\label{ln:op:expand}
  \EndWhile
  \State return $\Nk(q_t,p_t)$; $\Brem\gets\Brem-\spend_t$\label{ln:op:answer}
  \State \textbf{if} $s_t$: update $\hatC_c$ with $\Cost(q_t,p)$, $p\le p^\dagger$ \Comment{cost}\label{ln:op:cost}
  \State \textbf{if} $a_t=1$: update $\Rraw_c$ \Comment{replay audit}\label{ln:op:recall}
  \State $\lambda_{t+1}\gets\min\{\lammax,\max\{0,\lambda_t+\stepsize(\spend_t-b)\}\}$\label{ln:op:tariff}
\EndFor
\end{algorithmic}
\end{algorithm}\vspace{-2mm}

\subsection{Guarantees}\label{subsec:op:guarantees}
Let $\Rtrue_c(p)=\E[\Rec(q,p)\mid c(q)=c]$ and $\Ctrue_c(p)=\E[\Cost(q,p)\mid c(q)=c]$ be the true row of ball $c$, $\psi_c=\Prob[c(q)=c]$ its share of the stream, and $\omega_c=\E[1-\Rec(q,p_G)\mid c(q)=c]$ the miss rate of $p_G$ in $c$, so that $\sum_c\psi_c\omega_c=\bar\omega$. Regret is measured against the hindsight optimum $\OPT$, the largest total recall of any price assignment to the realized stream within $B$, which pays the same descent cost $\cpr(q)$ on every query as \oprobe, since no policy can price a query before probing it. Pricing a query from its ball rather than from the query itself costs the \emph{resolution gap} $\resgap\ge0$, the recall per query that per-ball prices lose against per-query prices in the fluid relaxation. It is zero when the queries of each ball share one row, and does not grow as balls are refined. The \emph{table errors} $\zeta^R_c(t)$ and $\zeta^C_c(t)$ bound, with probability at least $1-\delta$, how far the pooled rows $\Rraw_c$ and $\hatC_c$ of ball $c$ are from $\Rtrue_c$ and $\Ctrue_c$ when $q_t$ arrives. They fall as $O(\sqrt{\log(|\mathcal{B}|GN/\delta)/m})$ in the number $m$ of audits or cost records of $c$, with $N=|\Qa|+T$, and $\zeta^R_c$ adds $\omega_c$. They are stated for $\Rraw_c$ (the row \oprobe prices on) since cost-concave lifts can stay far from $\Rtrue_c$ irrespective of audits.

\begin{pbox}
\begin{theorem}[Regret]
Let $q_1,\dots,q_T$ be drawn i.i.d.\ from $\PP$, let Assumption~\ref{asm:feasible} hold, let $\Qa$ be drawn from $\PP$ independently of the tree, and let $\stepsize=\Theta(1/\sqrt T)$, $\lammax\ge1/(b-\floorbound)$ and $\delta=1/T$. Then
\[
\begin{aligned}
&\E[\OPT]-\E\big[\textstyle\sum_{t}\Rec(q_t,p_t)\big]\le T\resgap+\arate T\E[\Cost(q,p_G)]/\resrate\\
&\quad+O(\sqrt T)+2\textstyle\sum_{t}\E\big[\zeta^R_{c(q_t)}(t)+\lambda_t\zeta^C_{c(q_t)}(t)\big].
\end{aligned}
\]
\label{thm:regret}
\end{theorem}
\end{pbox}


\begin{pbox}
\begin{corollary}[Audit Rate]\label{cor:arate}
With $\psi_{\min}=\min_{c:\psi_c>0}\psi_c$ , $\arate=\min\{1,(T\psi_{\min})^{-1/3}\}$ and $b\le\caudmax$, the regret of Theorem~\ref{thm:regret} is $\widetilde O\big((1+\lammax\caudmax)\,T^{2/3}\psi_{\min}^{-1/3}\big)+T\resgap+2T\bar\omega$.
\end{corollary}
\end{pbox}

\noindent The bound separates two kinds of loss. The \emph{approximation loss} $T\resgap+2T\bar\omega$ is set by the balls and price list, not by \oprobe, and is linear in $T$ unless both gaps vanish. The rest is the \emph{learning overhead}, which is sublinear, so the result bounds the overhead above the approximation loss rather than giving sublinear regret against $\OPT$ in general. With exact tables ($\arate=0$, $\zeta^R=\zeta^C=0$) the overhead reduces to the $O(\sqrt T)$ of~\cite{balseiro2023dual}. The $T^{2/3}$ term is the cost of learning recall online. At the horizons of Section~\ref{sec:experiments} the $O(\sqrt T)$ term is of order of $T$, so our theorem is a statement of rates. The horizon $T$ is known, as in~\cite{balseiro2023dual}, since reserve check and $B=bT$ use it.
\section{$\textsc{Ledger}$ For The Batch}\label{sec:census}

\bprobe solves Problem~\ref{prob:batch} in three stages. \underline{First}: it probes every query of the batch, that is, serves it at the base price ($p_1$) and pauses its walk there (Section~\ref{subsec:probe}). \underline{Second}, it learns its tables from the audited queries of the training dataset, as described in the previous sections. \underline{Third}, finally it spends the rest of the budget on resuming the walks either exactly, greedily, or by using our anytime solution (Section~\ref{subsec:state}). Note that in the batched setting, the batch is known in full, so the allocation can be solved at once rather than \lq tracking\rq it, as we do for the online case.

\textbf{\underline{The Budget Allocation}}: with the tables of the \cube known \textit{exactly}, the batch allocation can be mapped to an instance of the multiple-choice knapsack problem~\cite{sinha1979multiple,kellerer2004knapsack}. Every query is allocated exactly one price from the list, and the search costs at that price must add up to at most the budget left after the probes and audits.

\textbf{\underline{\lbrute}}: the simplest way is to try every assignment of prices to queries and keep the best one whose predicted cost fits. It is exact, but does not scale beyond a handful of queries.

\textbf{\underline{\lexact}}: when the rows are the true rows of the queries, whose recalls are multiples of 1/k, the textbook dynamic program solution solves the problem exactly in $O(m^2Gk)$ time~\cite{kellerer2004knapsack}. 
This is polynomial in m and G and pseudo-polynomial in k. 

\textbf{\underline{\lgreedy}}: the greedy rule starts every query at $p_1$. It repeatedly takes the query whose next price in the list buys the most recall per unit of cost, applies that step, and stops when no further step fits in the budget 
For a query in ball $c$ at price $p_g$, the \emph{upgrade} moves it to $p_{g+1}$, and its \emph{marginal gain} is simply the recall that step buys per unit of cost,
\begin{equation}\label{eq:density}
\frac{R_c(p_{g+1})-R_c(p_g)}{\hatC_c(p_{g+1})-\hatC_c(p_g)} .
\end{equation}

Note that prices of equal predicted cost have equal recall on the majorant, so the step between them is free and is taken together with the next one, which basically keeps every gain defined. With rows that are cost-concave and nondecreasing (as we have discussed), and assuming the values in them are known exactly, the greedy is known to be within one upgrade of the optimum ~\cite{sinha1979multiple,kellerer2004knapsack}. The greedy runs in $O(mG\log m)$ time, and the searches then run in increasing order of assigned price (Section~\ref{sec:machinery}). A search stops before any expansion that could take the real spend of the batch above B, and the query then answers with the k closest vectors found so far. Any budget left goes to further upgrades in the same order.

\textbf{\underline{\lfptas}}: for any approximation factor $\varepsilon>0$ that a user picks, the textbook FPTAS (fully polynomial-time approximation scheme) solution~\cite{ibarra1975fast,lawler1979fast,kellerer2004knapsack} to the multiple-choice knapsack provides a budget allocation strategy within a factor $1-\varepsilon$ of the optimum in $O(m^2G/\varepsilon)$ time. As such, the FPTAS solution for the multiple-choice knapsack problem is well-known, and we point the reader to prior literature for more details~\cite{ibarra1975fast,lawler1979fast,kellerer2004knapsack}.

\begin{pbox}
\begin{proposition}[Known Tables]\label{thm:census}
Assume the rows of the tables of the \cube are the true rows of the queries of the batch, and let $B'$ be the budget left after the probes and the audits. Then \lexact returns an optimal allocation of Problem~\ref{prob:batch} at budget $B'$, \lfptas one within a factor $1-\varepsilon$ of it, and \lgreedy one within one upgrade of it, based on known guarantees~\cite{sinha1979multiple,kellerer2004knapsack,lawler1979fast}.
\end{proposition}
\end{pbox}

Note that these are the guarantees of the multiple-choice knapsack (Section~\ref{sec:model}). 

\subsection{$\textsc{AnyLedger}$: Serving Batches Anytime}\label{subsec:state}

\lstate is an adaptive budget allocator that monitors~\cite{hansen2001monitoring} the whole batch under the budget. It looks at the current state of the paused walks of queries, based on which it predicts what the next price (i.e., a higher beam) would buy for this query. Based on this prediction, it moves it up one price, and then reads the walk again, and repeats, and thereby completes every query in one heap.

\textbf{\underline{The State}}: the \emph{state} ($s$) of a paused walk is based of seven features that help predict whether the search can still be rewarding.

\vspace{-2mm}\begin{table}[H]
\centering\scriptsize\setlength{\tabcolsep}{3pt}
\caption{State Features Of The \lstate Walk}\label{tab:state}
\begin{tabular}{lll}
State & Definition & Meaning\\
\hline
new answers & \# of the $k$ answers new since the price below & still finding\\
shrink & $d_k$ over its value at $p_1$ & answers getting closer\\
frontier & next candidate's distance $/\,d_k$ & close: more to find\\
staleness & share of visits after the last answer found & time without progress\\
\hline
LID~\cite{levina2004mle,amsaleg2015estimating} & $\big(\mathrm{mean}_{i<k}\ln\frac{d_k}{d_i}\big)^{-1}$ & intrinsic dimension\\
$\epsilon$-hardness~\cite{ceccarello2025workloads} & \# of visited vectors within $1.1\,d_k$ & crowding\\
pool width & $d_{p_j}/d_k$ & width of the beam's pool\\
\end{tabular}
\end{table}\vspace{-2mm}

\textbf{\underline{The Budget Predictor}}: for every audited query $q'$ and price $p_j$, the replay of its audit gives its feature point, which we refer to its \lq state\rq\,at $p_j$ and its row above $p_j$, which tells us the recall gained and the cost paid to every higher price. A query $q$ at price $p_j$ is placed among them by rank. Every coordinate of its features and its state is read as the fraction of audited values at $p_j$ below it, so that no coordinate dominates it, and its row is the mean row of the $\kappa$ audited queries nearest to it in that rank space. The cost side is read as a ratio, the mean $\hat\ell$ over those $\kappa$ of $\log(1+\Cost(q',p_h))-\log(1+\Cost(q',p_j))$ which is the logarithm of the geometric mean of their cost ratios. A ratio is used because a query that has cost less or more than its neighbors so far tends to stay so in proportion. This is applied to the cost the walk has actually spent ($C_j$), which corrects a query that turned out cheaper or dearer than its neighbors. Then, the predicted cost at $p_h$ is simply computed as $(1+C_j)\,e^{\hat\ell}-1$. The predicted row is made nondecreasing over predicted cost (Figure~\ref{fig:concave}).

\textbf{\underline{Anytime Budget Allocation}}: remember that in the \probe, every query is probed. The heap orders the queries by their best ratio of the largest over $h>j$ of the predicted gain to $p_h$ over the predicted cost to $p_h$. Then the procedure works as follows. First, (a) the query at the top moves up one price, (b) its state is read again, (c) its row is predicted again, and then (d) it returns to the heap. This repeats until no walk can move or the cut rule stops a walk, which ends the batch. Our one step at a time approach lets the walk's own state confirm or deny a predicted gain before more is spent on it. Every move is a resumption under the cut rule, so the budget holds as for \lgreedy. 
\vspace{-2mm}\begin{algorithm}[H]\small
\caption{\lstate$(\mathcal{W},B)$}\label{alg:anytime}
\begin{algorithmic}[1]
\Require workload $\mathcal{W}$, budget $B$, price list $\mathcal{P}$, audited queries $\Qa$ of the training dataset, neighbors $K$, bound $\cex$
\State probe every $q\in\mathcal{W}$; $j(q)\gets1$ \Comment{Sec.~\ref{subsec:probe}}\label{ln:an:probe}
\State $H\gets\{(q,\hat\gamma(q))\}_{q\in\mathcal{W}}$ \Comment{max-heap on best ratio}\label{ln:an:heap}
\While{$H\neq\emptyset$}
  \State $q\gets\mathrm{pop}(H)$; $p^\dagger\gets p_{j(q)+1}$ \Comment{one price up}\label{ln:an:pop}
  \State resume $\Search(q,k,p^\dagger)$ while $\sum_{q'\in\mathcal{W}}\spend(q')+\cex\le B$\label{ln:an:resume}
  \State \textbf{if} $p^\dagger$ not reached: \textbf{break} \Comment{cut rule}\label{ln:an:cut}
  \State $j(q)\gets j(q)+1$
  \State \textbf{if} $j(q)<G$, candidates left: push $(q,\hat\gamma(q))$ \Comment{state re-read}\label{ln:an:push}
\EndWhile
\State return $\Nk(q,p_{j(q)})$ for $q\in\mathcal{W}$ \Comment{cut walk: what it holds}\label{ln:an:answer}
\end{algorithmic}
\end{algorithm}\vspace{-2mm}

\vspace{-2mm}\subsection{Optimizations}\label{sec:opt}

\underline{\textbf{Implemented:}} eight optimizations, besides the resumption and the replay of Section~\ref{sec:machinery}, we have: (a) \underline{Cost updates}: cost needs no audit, so every served query updates the cost row of its table. (b) \underline{Warm streams and batches}: a stream after a batch starts from the batch's audits and tariff, and a batch after a stream from the stream's. (c) \underline{Learning while idle}: between batches, \probe audits queries of the latest batch in the balls with the fewest audits, at no charge. (d) \underline{Parallel audits}: the audits are independent searches, run on all cores. (e) \underline{Solvers in parallel}: the knapsack solvers run the rows of tables on all cores in the serial order (the allocation unchanged). (f) \underline{8-bit code test}: a candidate whose distance on the library's 8-bit scalar codes~\cite{douze2024faiss} clears the walk's threshold by more than the quantization bound is dropped without its exact distance, the deterministic form of the cheap distance comparisons of \textsc{ADSampling}, \textsc{RaBitQ} and \textsc{Finger}~\cite{gao2023adsampling,gao2024rabitq,chen2023finger}, so every answer is unchanged and a code distance is charged half a distance computation. (g) \underline{Restart}: \lstate may search a query that has settled away from its neighbors once more, at the base price from a random vertex of an upper layer, and merge the answers into its paused walk, when the predicted gain per cost of this restart beats the next price, in the spirit of the restart policies of~\cite{kautz2002restart}. (h) \underline{Reuse}: in the spirit of \textsc{Aker}~\cite{oh2026aker}, every audited query's top-price answer is cached under the layer-one vertex its probe reached, and a new query pays one distance per entry under its own vertex, takes the nearest, and seeds its walk with that entry's $k$ answers (charged from budget).

\section{Experimental Evaluation}\label{sec:experiments}
\providecommand{\TBD}{{\color{red}\textbf{?}}}

This section evaluates \probe. In our study, we use distance computations as a measure of budgets, and report recall against the exact neighbors, so that our numbers are easy to compare with. \probe is implemented on HNSW in C++ on \texttt{FAISS} and runs on an Intel Core i7-13700H with 14 cores, 64\,GB of memory and Ubuntu 26.04. We use the datasets of Table~\ref{tab:datasets} and features described below.
\newline\indent\textbf{\underline{Features}}: we remind the reader that the design of \probe is inherently feature-agnostic, and any of the heuristic features defined in prior works~\cite{learnedtermination3,ceccarello2025workloads,tao,darth2025,quake,auncel,adaptivebeam} can be plugged in to our framework. Nevertheless, based on calibrations performed on our datasets, we present and use the features in Table~\ref{tab:features} for our study. The \cube reads nine, and the predictor of \lstate reads three more together with the seven state values of Section~\ref{subsec:state}. Table~\ref{tab:datasets} lists our datasets, which differ in dimensionality and modality.

\vspace{-2mm}\begin{table}[H]
\centering\scriptsize\setlength{\tabcolsep}{2pt}
\begin{minipage}[c]{69pt}
\centering
\caption{Datasets}\label{tab:datasets}
\begin{tabular}{@{}lcc@{}}
\toprule 
dataset & $n$ & $d$\\
\midrule 
DEEP-10M & $10^7$ & 96\\
SIFT-10M & $10^7$ & 128\\
SpaceV-10M & $10^7$ & 100\\
T2I-10M & $10^7$ & 200\\
\bottomrule
\end{tabular}
\end{minipage}\hfill
\begin{minipage}[c]{167pt}
\centering
\caption{Features (Last 3: \lstate)}\label{tab:features}
\begin{tabular}{@{}lll@{}}
\toprule
feature & definition & what it captures\\
\midrule 
scale & $D[k'-1]$ & radius to reach\\
spread & $D[k'-1]-D[0]$ & unevenness\\
median & $D[\lfloor k'/2\rfloor]$ & where the bulk sits\\
LID~\cite{levina2004mle,amsaleg2015estimating} & $\big(\mathrm{mean}_{i}\ln\frac{D[k'-1]}{D[i]}\big)^{-1}$ & intrinsic dimension\\
reach & \# vectors evaluated & descent length\\
entry dist. & entry point to $q$ & remoteness\\
degree & mean base degree of $D$ & connectivity\\
RC~\cite{he2012difficulty,ceccarello2025workloads} & $\bar D_q/D[k'-1]$ & relative contrast\\
expansion~\cite{aumuller2019role,ceccarello2025workloads} & $D_{\mathrm{all}}[2k'-1]/D[k'-1]$ & expansion\\
\midrule
$\epsilon$-hardness~\cite{ceccarello2025workloads} & \# $D_{\mathrm{all}}\le1.1\,D[k'-1]$ & crowding\\
in-degree & mean base in-degree of $D$ & hubness~\cite{radovanovic2010hubs}\\
exact RC & $\bar d_q/D[k'-1]$ & relative contrast\\
\bottomrule
\end{tabular}

\vspace{1pt}\raggedright $D$: the $k'$ nearest of the layer vectors the descent evaluated, sorted; $D_{\mathrm{all}}$: all of them; $\bar D_q$: their mean; $\bar d_q$: RMS distance to the collection.
\end{minipage}
\end{table}\vspace{-2mm}

\begin{figure*}[t]
\centering
\includegraphics[width=\textwidth]{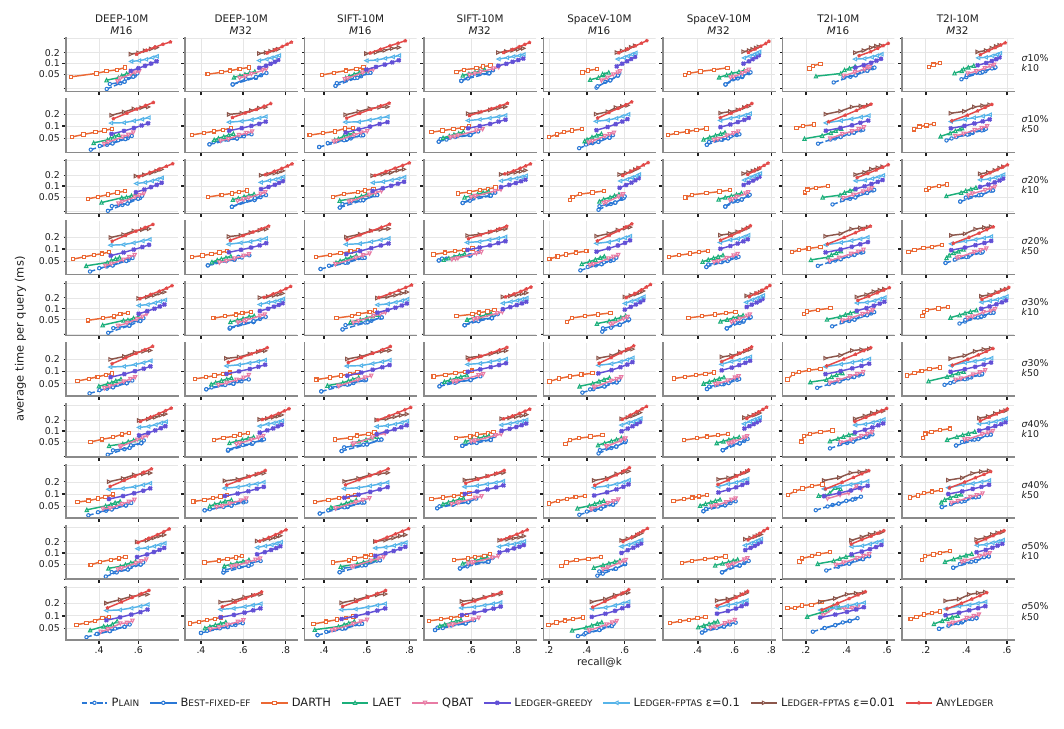}
\vspace{-2mm}\caption{Time Vs Recall; Every Budget Is A Point Along Each Approach's Curve}\vspace{-2mm}\label{fig:curves}
\end{figure*}

\textbf{\underline{Workloads}}: adding Gaussian noise to seed vectors is the standard and most established way of generating vector workloads~\cite{zoumpatianos2018generating,echihabi2022hercules,azizi2023elpis,azizi2025graph,darth2025,li2026reweave}. In the spirit of prior literature, we pick a few vectors as seeds, and every query of the workload is then one seed plus noise. Our seeds are real queries, and we call them \emph{reference queries}. For a diverse query workload, our reference queries are the $100$ real queries with the lowest recall under plain search at $\mathrm{ef}=40$ and the $100$ with the highest recall. Each workload has $10{,}000$ queries in total. Each query is a \emph{draw}, i.e., a reference query picked uniformly at random plus Gaussian noise. The noise has variance $\sigma^2 s^2/d$ per coordinate, where $d$ = dimension and $s$ = spread of reference queries (their mean distance to their centroid), so its norm is about $\sigma s$. We use $\sigma$ of 10, 20, 30, 40 and
50\%. The reference queries are split into four folds of equal size. Each fold serves once as the test fold. 

The test batch is every query drawn around the reference queries of that fold. The training dataset is every query drawn around the reference queries of the other three folds, so a test query never shares its reference query with a training query. We use \textsc{HNSW} under the Euclidean distance with $M=\{16,32\}$, $\mathrm{efConstruction}=100$ and $k=\{10,50\}$, and the price list is $\mathrm{ef}\in\{10,20,\dots,300\}$, so the base price is $\mathrm{ef}=10$. We set $r=8$, $k'_{\min}=8$, $m_{\max}=50$ and $\kappa=300$. If a price is below $k$, the search keeps $k$ candidates from the heap. Every batch is run at six budgets. The smallest is the \emph{base budget}, the number of distance computations the plain search of the library needs to serve every query of the batch at the base price, rounded up to the next thousand. The other five add $100$, $200$, $300$, $400$ and $500$ distance computations per query to it. Every variant of the \probe runs with the 8-bit code test of Section~\ref{sec:opt}, in the batch experiments also with its reuse and \lstate with its restart, and a code distance counts half a distance computation in its budgets.

\textbf{\underline{Baselines}}: we consider the following. (a) \underline{\textsc{Plain}} is HNSW search at the one ef whose cost on the training dataset fits the budget, what a user without \probe runs. (b) \underline{\textsc{Best-fixed-ef}} is the best single ef at each budget, chosen in hindsight on the test queries, an upper bound on the static setting. A query its budget does not reach scores zero. (c) The \underline{target-based methods} \textsc{DARTH}~\cite{darth2025}, \textsc{LAET}~\cite{learnedtermination3} and \textsc{QBAT}~\cite{qbat2026} take a recall target, not a budget, and decide each query alone. We sweep the target from $0.01$ to $0.99$ in steps of $0.01$ on the training dataset, keep the one whose spend there is the closest to the budget from below, and report the recall the method then reaches on the test batch, \underline{\textbf{at matched budget spend}}. If its spend on the test batch is above the budget, we step down to the next lower target, so no reported value is over its budget. This is the best such a method can do under a budget. All three are ported to the same index and search under a beam of $100$ candidates. A method is reported only at the budgets that its lowest target fits on the training dataset. (d) \underline{\lgreedy}, \lexact (rounded to $10^{-3}$), \underline{\lfptas} with $\varepsilon$ of $0.1$ and of $0.01$, and \underline{\lstate} have been discussed earlier. (e) \underline{\textsc{Brute-force}} is the exact program on the true rows of the test queries, the most any allocation of these prices can reach.

\begin{figure*}[t]
\centering
\includegraphics[width=0.9\textwidth]{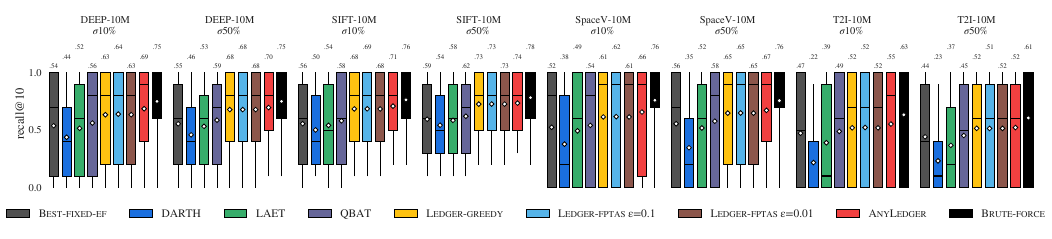}
\vspace{-2mm}\caption{Per-Query Recall Spread At $\sigma$ Of $10\%$ And $50\%$, $M{=}16$, $k{=}10$, $+300$/Query}\label{fig:sigma}
\end{figure*}\vspace{-2mm}

\begin{figure*}[t]
\centering
\includegraphics[width=0.9\textwidth]{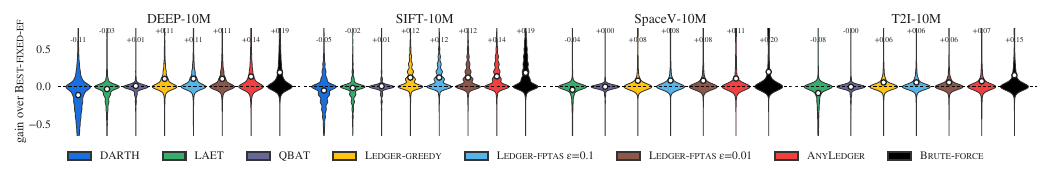}
\caption{Per-Query Gain Over \textsc{Best-fixed-ef} At Equal Budget, Pooled Over $\sigma$ And The Five Budgets, $M{=}16$, $k{=}10$}\label{fig:gains}
\end{figure*}

\subsection{Results For Batched Workloads}\label{subsec:exp-batch}

\textbf{\underline{Allocation Performance}}: Figure~\ref{fig:curves} evaluates \probe across different workloads, and reports the average time per query against recall. Notably, every budget is a point in Figure~\ref{fig:curves} along the curve of each approach. Our results show that at equal budget \lgreedy beats both \textsc{Plain} and \textsc{Best-fixed-ef} that most practitioners use, by about eight points of recall over \textsc{Best-fixed-ef} on average. \lstate is above all three on every workload over the four folds, and above \lgreedy by $1.5$ points on average. The gain of \lstate over \lgreedy\ falls as $\sigma$ grows. It is largest at $\sigma=10\%$, where the easy and the hard regions are sharpest, and at $\sigma=50\%$ it is lower. The gain of every allocation over \textsc{Best-fixed-ef} stays about level across the noise levels, between seven and ten points. In general, we are able to see that \lstate and the other \probe-based approaches offer significant gains over recall target-based methods, due to their budget-aware nature.
 
\textbf{\underline{Per-Query Recall}}: Figure~\ref{fig:sigma} presents the spreads of the per-query recall of every approach at our default settings, and Figure~\ref{fig:gains} the per-query gain over \textsc{Best-fixed-ef} at the same budget. The gains are pooled over the noise levels and the five budgets, with the mean and its interval over the folds on every plot. Our results show that the allocation does not buy high mean recall by starving hard queries. Instead, the share of queries below half recall falls at every step, from $43\%$ under \textsc{Best-fixed-ef} to $33\%$ under \lgreedy, $29\%$ under \lstate and $22\%$ under \textsc{Brute-force}. The fifth percentile of the per-query recall rises with it, from $0.01$ under \textsc{Best-fixed-ef} to $0.05$ under \lgreedy and $0.08$ under \lstate. As $\sigma$ grows, the boxes rise and widen, since the draws wander off the hard spots, and the gap between \lstate and \lgreedy\ narrows with them. Our results show that the gains over existing baselines are significant, across multiple configurations and heterogeneous datasets.

\textbf{\underline{Comparison To Brute Force}}: \textsc{Brute-force} performs better than \lstate on every workload, and stays above by about six points of recall on average. This is because the rows are merely estimates of cost-recall growth values with price, and a prediction that is wrong can move a query one price too early or too late.

\begin{figure*}[t]
\centering
\includegraphics[width=\textwidth]{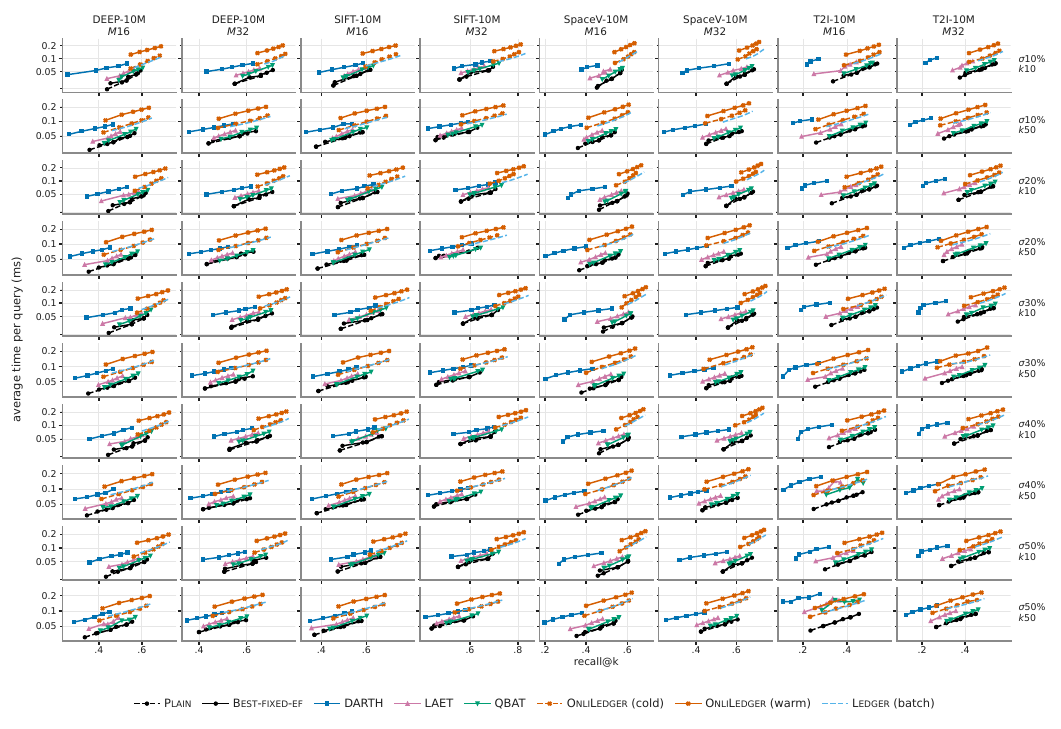}
\caption{Time Against Recall On The Stream; Every Budget Rate Is A Point Along Each Approach's Curve}\label{fig:streamcurves}
\end{figure*}

\textbf{\underline{Time}}: our runs also report their serving time per query, which is the allocation (the knapsack or the predictor) plus the probe and the search. Every time is measured on one thread, and \lexact and \lfptas alone run on eight. Our results indicate that \lgreedy returns the allocation extremely fast with the tree and its tables, so its allocation adds little to the serving time. \lstate is slower per query as \lgreedy, since it reads every walk again after every move, and its predictions cost more than its searches. 

\lexact takes about $0.9$ milliseconds per query on eight cores. The three solvers are the same framework at three price points, \lgreedy when the allocation must cost nothing, \lexact when the batch is worth an exact answer on its tables, \lstate when the walks' own progress is worth reading. All three improve on the fixed prices of the state of the art at equal budget, and we leave the choice to the application. Our approaches show huge improvements over the baselines while staying within budget, and the extra time incurred is a small cost to pay for it. For the same recall \lgreedy therefore takes $1.4$ to $2.1$ times the time of \textsc{Plain} at a larger ef on our datasets and \lstate $3.3$ to $4.1$ times. So \probe pays where a distance computation is the expensive step, as in the settings of Section~\ref{sec:intro}, and not where the vectors sit in memory and time alone is counted.

\subsection{Results For Online Workloads}\label{subsec:exp-stream}

\begin{figure}[H]
\centering
\includegraphics[width=\columnwidth]{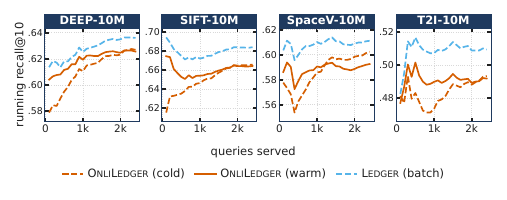}\vspace{-2mm}
\caption{Running Recall Along Streams At $+300$/Query}\label{fig:stream}
\end{figure}

Here the queries arrive one at a time in a random order, and \onli\ serves them under a budget rate, which is simply set as the budget per query of the batch experiments. It sets $\arate=0.1\,b/\caudmax$, $\lammax=1/(b-\floorbound)$ and $\stepsize=\lammax/(\caudmax\sqrt T)$, with the bounds $\floorbound$ and $\caudmax$ set from the training dataset and the rate. Every stream is the test fold of a workload in a random order, about $2{,}500$ queries, and starts from the same training dataset as the batch run of that fold. \onli\ runs warm, with its tables filled from the audited queries of the training dataset (about $7{,}500$), and cold, with empty tables (the tree, the bounds and the first tariff still come from the training dataset). \cart on the same queries is the batch allocation a stream can hope to match. Figure~\ref{fig:streamcurves} shows the time-against-recall view of the stream (every rate is a point). We see that \textsc{Plain}, \textsc{Best-fixed-ef} and the target-based methods decide each query alone, so a stream is the same as a batch to them, and their curves are the batch runs on the same queries. \onli\ is the only method that tracks a budget rate online. The warm stream starts from the tables of the training dataset. The cold stream starts at the highest price the rate affords for every query, since the root row holds the only audits, and climbs as balls detach from the root one by one. Figure~\ref{fig:stream} follows the recall of a stream as its queries arrive ($\sigma=10\%$, $M=16$, $k=10$).

\underline{\textbf{Summary:}} (a)~at equal budget every method of \probe gains over \textsc{Best-fixed-ef}, \lgreedy by about eight points of recall, and \lstate, ahead on every workload, adds $1.5$ points over it, (b)~the gain of \lstate is largest where the workload is most concentrated, (c)~the state of the walk, read after every upgrade, closes part of the gap to \textsc{Brute-force}, which stays well above every method, (d)~the allocation costs microseconds to a fraction of a millisecond per query and the figures charge it, while in time alone \textsc{Plain} stays ahead, and (e)~every budgeted method holds its budget, by the cut rule, where a target-based method's spend is its own.

\vspace{-3mm}\vspace{-3mm}
\section{Related Work And Conclusion}\label{sec:related}

\underline{Graph Indexes And Effort Tuning:} proximity graphs rule over the state-of-art in ANN search~\cite{wang2021comprehensive,malkov2020hnsw,fu2019nsg,subramanya2019diskann}, and most if not all vector databases expose the search beam as a per-query parameter of the search call~\cite{douze2024faiss,wang2021milvus,pgvector}. \probe treats the index as a black box.
\newline\indent\underline{Adaptive Effort And Early Termination:} queries differ in the effort they need, and every method that adapts it takes recall as the input \textsc{Laet}~\cite{learnedtermination3} and \textsc{Tao}~\cite{tao} predict the remaining work, \textsc{Darth}~\cite{darth2025}, \textsc{Vader}~\cite{chatzakis2026vaderfilteredvectorsearch}, and \textsc{Quake}~\cite{quake} adapt effort to a target, \textsc{ANNiE}~\cite{annie2026}, \textsc{QASP}~\cite{qasp2026} and \textsc{Sheaf}~\cite{sheaf2026} predict a query's cost or recall curve, and others bound the error~\cite{auncel,adaptivebeam,hao2026recallaware} or repair a finished search~\cite{li2026spanner}. All decide each query alone. \probe inverts the input: the budget is given, recall is bought, and the saving on an easy query is spent on a hard one. Top performing and latest works from the above list constitute our baselines in this work.
\newline\indent\underline{Query Hardness And Budgeted Processing:} hardness can be measured in many ways, including: offline by relative contrast~\cite{he2012difficulty}, expansion~\cite{aumuller2019role}, local intrinsic dimensionality~\cite{amsaleg2015estimating} and Steiner hardness~\cite{wang2024steiner,ceccarello2025workloads}, and any of them can join the features of the \cube. Spending a budget rather than reaching a target is old in information retrieval~\cite{zilberstein1996,cambazoglu2010early,mackenzie2022anytime,broccolo2013load}. In vector search a budget has traditionally meant a per-query cap~\cite{baranchuk2019routing,ram2012timebudget}. To our knowledge, no method in vector search moves effort \emph{between} the queries of the batch it serves by their marginal recall gains under a hard budget. The same recipe is known in web-search scheduling~\cite{he2012zeta}, recommender serving~\cite{jiang2020dcaf}, budgeted classification~\cite{huang2018msdnet} and LLM inference~\cite{damani2025hard}, with independent requests and nothing read from a running one.
\newline\indent\underline{Budgets, Learning And Anytime Algorithms:} a budget with a knapsack behind it is old in databases, in index selection~\cite{chaudhuri1997efficient}, approximate query processing~\cite{agarwal2013blinkdb}, buffer allocation by marginal gain~\cite{ng1991flexible} and load shedding~\cite{babcock2004load}, and the \cube is a novel learned structure in works on learned indexes~\cite{kraska2018case}, with a model of recall against cost in place of a fixed effort. The monitoring of anytime algorithms~\cite{svegliato2018meta}, anytime nearest neighbor queries~\cite{ciaccia2000pac,xu2008anytime} and budget-limited bandits~\cite{tranthanh2012knapsack} are ancestors of \lstate. Online knapsacks and budget pacing give the dual form \onli uses~\cite{devanur2009adwords,balseiro2023dual}.

\textbf{\underline{Conclusion And Future Work:}} this paper formally introduces and addresses the problem of budgeted ANN search over a query workload (streaming and batched), with the goal to maximize total recall. Our key idea is to read the hardness of a query from the sparse layer that its search at the base price passes, which we call its probe, and to populate the tables of \cube from it. \probe uses these tables to quickly estimate cost-recall growth across prices, and distributes the budget strategically across the workload. Our future efforts will look at (a) dynamic indexes with insertions and deletions, (b) scenarios where budgets and recall targets are traded off on the fly, and (c) integration of \probe into existing systems.\vspace{-2mm}

\bibliographystyle{ACM-Reference-Format}
\bibliography{sample}


\end{document}